\documentclass[10pt,journal,compsoc]{IEEEtran}
\newtheorem{theorem}{Theorem}
\newtheorem{definition}{Definition}

\newcommand{\tabincell}[2]{\begin{tabular}{@{}#1@{}}#2\end{tabular}}

\usepackage{verbatim}
\usepackage{graphicx,color}
\usepackage{algorithm}
\usepackage{algorithmic}
\usepackage{rotating}

\usepackage{amsmath, amssymb}
\usepackage{breqn}
\usepackage{threeparttable}

\usepackage{multirow}
\usepackage[numbers,sort&compress]{natbib}
\usepackage{subfig}
\usepackage{url}
\usepackage{hyperref}
\usepackage{breakurl}
\usepackage{booktabs}
\usepackage{color}
\usepackage{color}
\usepackage{framed}
\definecolor{shadecolor}{gray}{0.85}
\usepackage{makecell}

\ifCLASSOPTIONcompsoc
  \usepackage[nocompress]{cite}
\else
  \usepackage{cite}
\fi
\ifCLASSINFOpdf
\else
\fi
\begin{document}
 \pdfoutput=1 ！！！！
\title{Local Differential Privacy and Its Applications: A Comprehensive Survey}
%
%
%
%

\author{Mengmeng Yang,~Lingjuan Lyu,~Jun Zhao,~Tianqing Zhu,~Kwok-Yan Lam

\IEEEcompsocitemizethanks{\IEEEcompsocthanksitem M. Yang, J. Zhao, and K. Y. Lam are with Nanyang Technological University, Singapore. \protect\\
E-mail: $\{$melody.yang, junzhao, and  kwokyan.lam$\}$@ntu.edu.sg
\IEEEcompsocthanksitem Lingjuan Lyu is with National University of Singapore. \protect\\
E-mail: dcslyul@nus.edu.sg
\IEEEcompsocthanksitem Tianqing Zhu is with University of Technology Sydney, Australia. \protect\\
E-mail: Tianqing.Zhu@uts.edu.au}
\thanks{Manuscript received April 19, 2005; revised August 26, 2015.}}

%
%

\markboth{Journal of \LaTeX\ Class Files,~Vol.~14, No.~8, August~2015}%
{Shell \MakeLowercase{\textit{et al.}}: Bare Demo of IEEEtran.cls for Computer Society Journals}
%



\IEEEtitleabstractindextext{%
\begin{abstract}
With the fast development of Information Technology, a tremendous amount of data have been generated and collected for research and analysis purposes. 
As an increasing number of users are growing concerned about their personal information, 
privacy preservation has become an urgent problem to be solved and has attracted 
significant attention. 
Local differential privacy (LDP), as a strong privacy 
tool, has been widely deployed in the real world in recent years. 
It breaks the shackles of the trusted third party, and allows users to perturb their data 
locally, thus providing much stronger privacy protection. 
This survey provides a comprehensive and structured overview of the local differential privacy technology. 
We summarise and analyze state-of-the-art research in LDP and compare 
a range of methods in the context of answering a variety of queries and training different machine learning models. 
We discuss the practical deployment of local differential privacy and explore its application in various domains. 
Furthermore, we point out several research gaps, and 
discuss promising future research directions.

\end{abstract}

}

\maketitle

\IEEEdisplaynontitleabstractindextext

%
\IEEEpeerreviewmaketitle

\IEEEraisesectionheading{\section{Introduction}\label{sec:introduction}}
Due to the promise of data-driven decision-making is now being recognized broadly, the collection of user data has increased dramatically over the last few years. 
However, the 
collected information from users is generally private and sensitive, 
which can be easily linked to other highly confidential details. 
Users raise the privacy concern over their personal data, especially after the emergence of the new technology for the in-depth mining and analysis of the users' data. 
Various countries have enacted privacy laws to regulate the actives of the organization with users' data, 
such as General Data Protection Regulation (GDPR) \citep{GDPR}, California Consumer Privacy Act (CCPA) \citep{CCPA} and Personal Data Protection Act (PDPA) \citep{PDPA}. 
Privacy preservation, therefore, becomes an urgent issue that needs to be addressed.

Differential privacy 
was proposed by Dwork et al. \citep{dwork2006our}, and 
has emerged as an \textit{de facto} standard for preserving privacy in a variety of areas. 
The traditional differential privacy, also named centralized differential privacy (CDP), collects the user's original data first and then releases the perturbed aggregated information to the public user. 
It assumes the data curator is trusted, which is not always the case in the real world. 
Even the big reputable company cannot guarantee its customer's privacy. 
For example, Google exposed the private data of hundreds of thousands of users of the Google+ social network in 2018. 
In 2019, more than 267 million Facebook users had their user IDs, phone numbers, and names exposed online \citep{fb}. Reported in Feb, 2020, personal data of all 6.5 million Israeli voters is exposed because of the elector app flaw, including full names, addresses, and identity card numbers, which raises concerns about identity theft and electoral manipulation \citep{pdamiv}.  
Therefore, it is difficult to find a fully trusted third party to manage the users' data. 

\begin{figure}[tbp]
\centering

\subfloat[Centralized differential privacy ]{
\label{Fig-Gowalla01}
\includegraphics[scale=0.5]{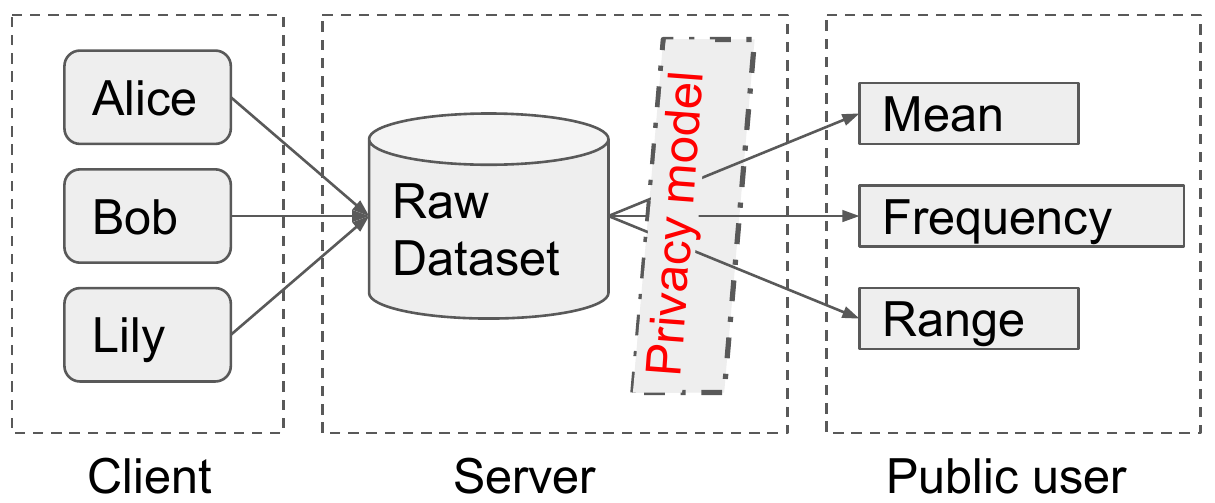}
}

\subfloat[Local differential privacy ]{
\label{Fig-knnmGres}
\includegraphics[scale=0.5]{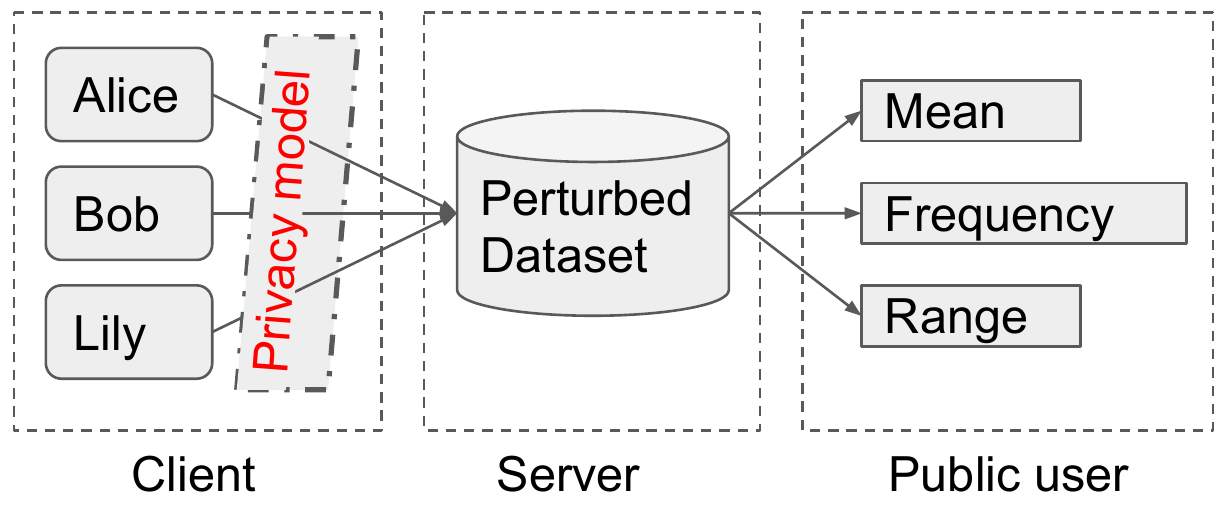}
}

\caption{Comparison of two differential privacy models}
\label{cdpm}
\end{figure}

To solve this problem, local differential privacy (LDP) \citep{kasiviswanathan2011can} was proposed. It perturbs the user's data locally before the data leaves the users' devices. Only the data owner can access the original data, which provides much stronger privacy protection for the user. Fig. \ref{cdpm} shows the comparison of the framework of centralized differential privacy and local differential privacy. 
For centralized differential privacy, the data curator possess the users' true data. 
While under the local differential privacy model, the curator holds the perturbed data instead of the raw data, the query is performed on the perturbed dataset. 
Therefore, local differential privacy prevents the privacy disclosure from the untrusted data curator and relieves the burden on the trusted data curator to keep data secure. 

LDP received a significant amount of attention and has been deployed by many big companies \citep{erlingsson2014rappor,adp,ding2017collecting} to preserve users' privacy.  
However, it adds too much noise to the whole dataset, which results in much lower utility compared with central model and each user only has a local view of the data, which limits the scope of its application. 
Therefore, the local differential privacy has been much less studied than the central one. 
In order to summarize the state-of-the-art and to benefit future research, we are motived to provide a survey about current work in this field. 
Though there exist surveys on LDP, they only provide partial pictures. 
For example, Zhao et al. \citep{zhao2019survey} mainly investigate the potential applications of LDP in securing the internet of connected vehicles. Bebensee \citep{bebensee2019local} focuses on heavy hitter identification and spatial data collection. Cormode et al. \citep{cormode2018privacy} provide a brief tutorial on LDP. 
To the best of our knowledge, this is the first work that comprehensively studied the research over local differential privacy, including its technologies and applications.

Specifically, we identify two research directions according to the dependency of data perturbation mechanism: statistical query with LDP and private learning with LDP. 
For statistic query with LDP, the aggregator aims to collect users' data to answer a specific query, such as frequency, mean, and range. While the data perturbation mechanism is designed based on the specific query type. 
For private learning with LDP, the aggregator aims to collaborate with users to train a model, which is associated with particular machine learning algorithms, such as linear regression and SVM. 
The data perturbation mechanism is dependent on the specific machine learning algorithm.
We point out the research problem and identify the unit challenge for each statistic query and model training under local differential privacy protection. 
We list and review the related methods and techniques, and analyze their advantages and disadvantages. 
Besides, we discuss the practical deployment of LDP and explore the latest research progress over the development of LDP's applications, including federated learning, reinforcement learning, social network, location privacy, and recommendation system. 
Furthermore, according to the latest research and character of LDP, we identify the research gaps and discuss the research directions in the near further. 

This survey is structured as follows. In Section \ref{pre}, we introduce some preliminaries. The review of exiting LDP techniques for statistic query and machine learning are presented in Section \ref{sqldp} and Section \ref{mlldp} respectively. Section \ref{aldp} explored the application of LDP. Section \ref{rgrd} provides an extensive discussion on research gap and research directions for LDP research. We finally conclude this survey in Section \ref{con}.

\section{Preliminaries}\label{pre}
\subsection{Notations}
Let $\mathcal{V}$ be the user's record with $d$ attributes, which samples from a finite data universe $\mathcal{X}$. 
We consider a set of $n$ users, where each user $u_i$ has one or some items $v_j\in \mathcal{V}$. 
Each user needs to report his value to the aggregator for analysis purpose. 
Local differential privacy aims to hide the user's true value by perturbing the report while ensuring statistical accuracy. 
The local differential privacy is achieved by a mechanism $\mathcal{M}$, which is a randomized algorithm that perturbs the user's true value as a random value $\hat{t}$. 
The aggregator collects $\hat{t}$ and estimates the true statistics. 
Table \ref{tab-parameters} summarizes the notations used in the following sections.

\begin{table}[htpb]  \centering
  \caption{Notations}\label{tab-parameters}
  \begin{tabular}{clcl}
\hline
    Notation & Description& Notation & Description  \\
\hline
$\ell(\cdot)$ & Loss function & $n$ &The number of users\\
$v$ & User's value & $\Psi$ & Estimation algorithm\\
$t$ & Encoded value & $\mathbb{H}$ & Hash function Universe\\
$\hat{t}$ & Perturbed value & $p,q$ & Perturbation probability\\
$A$ & Attribute & $g$ & Number of segment\\
$I_i$ & Items of attribute & $C$ & Cartesian product\\ 
$c(I)$ & Count of $I$ &$\mathcal{M}, \Phi$ & Randomized algorithm \\
$r$ & User's report & $\mathbf{w}$ & Output model\\
$f$ & Frequency & $\alpha, \beta$ & Accuracy parameter\\
$\mathcal{H}$ & Hash function & $h$ & Hypothesis\\
$\epsilon$ & Privacy budget & $\delta$ & confidence parameter\\

\hline
\end{tabular}
\end{table}

\subsection{Local Differential privacy}

\begin{definition}[$\epsilon$-Local Differential Privacy \citep{duchi2013local}]  
A randomized algorithm $\mathcal{M}$ satisfies $\epsilon$-local differential privacy, if and only if for any pair of input values $v, v'\in D$ and for any possible output $S\subseteq Range(\mathcal{M})$, we have 
\begin{equation*}
Pr[\mathcal{M}(v)\in S] \leq e^\epsilon Pr[\mathcal{M}(v')\in S].
\end{equation*}
\end{definition}
The perturbation mechanism $\mathcal{M}$ is applied to each user record independently. 

The majority of local differential privacy mechanisms are based on the idea of randomized response \citep{1965}, which was proposed by Warner et al. as a survey technology to eliminate the evasive answer bias. 

\begin{definition}[Randomized Response (RR)] 
Let $v$ be a user's binary value, $\hat{t}$ be the response
Then, for any $v$,

\begin{equation}\label{LDPRR}
Pr[\hat{t}=v]=\begin{cases}
\frac{e^\epsilon}{e^\epsilon+1},& \mathrm{if} ~t=v\\
\frac{1}{e^\epsilon+1},& \mathrm{if} ~t\neq v
\end{cases}.
\end{equation} 
\end{definition}
 
The randomized response outputs the true value with probability $\frac{e^\epsilon}{e^\epsilon+1}$ and outputs the opposite value with probability $\frac{1}{e^\epsilon+1}$. 
Holohan et al. \citep{holohan2017optimal} show that the output probability of Eq. \ref{LDPRR} minimises estimator error. 
However, the traditional randomized response is only suitable for the binary attribute. 
To make it suitable for a larger domain, a \textit{generalized randomized response (GRR)} is proposed \citep{kairouz2016discrete}. 

\begin{definition}[Generalized Randomized Response (GRR)]
Given a user $u$ with a value $v\in R$, where $R$ is a set of $d$ possible true values that a user can have.   
A random variable, denote by $\hat{t}$, represents the response of a user $u$ with sample space $R.$ 
The generalized randomized response works as follow:

\begin{equation}
Pr[\hat{t}=v]=\begin{cases}
\frac{e^\epsilon}{e^\epsilon+d-1},& if ~t=v\\
\frac{1}{e^\epsilon+d-1},& if ~t\neq v
\end{cases}.
\end{equation} \label{PDRR}

\end{definition}
When the data dimension $d=2$, the generalized randomized response is consistent with the traditional randomized response.

Adding Laplace or Gaussian noise to the data record achieves LDP as well. 
It is more used for continuous numerical data statistics. 
The definition is shown as follows. 

\begin{definition}[Laplace mechanism]  
Given a function $f: D \rightarrow \mathbb{R}$ over a dataset $D$, 
the Laplace mechanism is defined as:
\begin{equation*}
\widehat{f}(D)=f(D)+Laplace(\frac{s}{\epsilon}) ,
\end{equation*}
\end{definition}

\begin{definition}[Gaussian mechanism]  
Given a function $f: D \rightarrow \mathbb{R}$ over a dataset $D$, 
the Gaussian mechanism is defined as:
\begin{equation*}
\widehat{f}(D)=f(D)+\mathcal{N}(u,\sigma^2s^2) ,
\end{equation*}
\end{definition}

where $s$ is the sensitivity. 
Usually, a clipping technique \citep{abadi2016deep} or truncation are adopted to bound the sensitivity to achieve a better performance.

Since the local differential privacy is based on the theory of differential privacy, the sequential composition property of differential privacy applies to local differential privacy. 

\begin{theorem}[Sequential Composition \citep{wang2018privtrie}]
Suppose a method includes $m$ independent randomized functions $\mathcal{M} =\{\mathcal{M}_1, \mathcal{M}_2, ..., \mathcal{M}_m \}$, each $\mathcal{M}_i$ provides $\epsilon_i$-local differential privacy guarantee, then $\mathcal{M}$ satisfies $(\sum_{i=1}^m \epsilon_i)$-local differential privacy. 
\end{theorem}

More sophisticated algorithms can be achieved based on the sequential composition property. 
Specifically, the aggregator can apply a series of local differential privacy mechanisms, assign a portion of privacy budget to each of them, the series of mechanisms as a whole satisfies $\epsilon$-local differential privacy.

\subsection{Data aggregation under LDP}
\begin{figure}[!htp]
\centering
\includegraphics[width=3.5in]{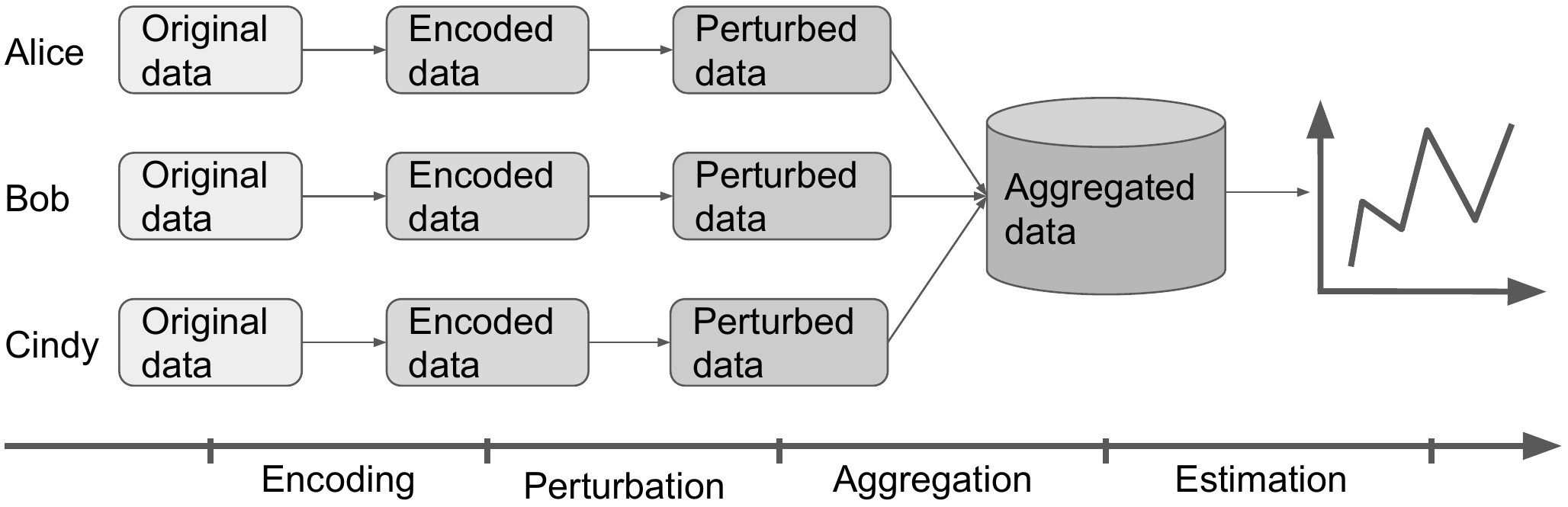}
\caption{Data flow of statistics under LDP}
\label{Model}
\end{figure}

LDP-perturbed value is randomized and no longer holds any significance, 
but the aggregates like average across large numbers of participants still do. 
Therefore, the application of LDP is based on the aggregation over a large volume of participants. 
Data aggregation under local differential privacy mainly includs four processes: Encoding, Perturbation, Aggregation, and Estimation. 
The majority of LDP algorithms follow this aggregation process, which is shown in Fig. \ref{Model}.

\begin{itemize}
\item \textit{Encoding.} 
The user encodes the original value $v$ to $t$ according to the predefined coding scheme, 
which makes it adapted to the perturbation mechanism. The encoded $t$ can be a value or a binary vector. 
\item \textit{Perturbation.} 
The encoded value $t$ is perturbed by the randomized algorithm that achieves local differential privacy, which guarantees that the user's value cannot be distinguished with any other values. 
\item \textit{Aggregation.} 
The aggregator collects all the perturbed value $\hat{t}$ from users and aggregates them accordingly. 
\item \textit{Estimation.} 
The aggregator estimates the query result according to the perturbation strategy to ensure statistical unbiased, and may utilize some post-processing techniques to improve estimation accuracy.  

\end{itemize}

\section{Statistical query with LDP} \label{sqldp}
Capture simple queries is a very useful primitive in building more complicated data analyses, which help service provider to better understand the needs of populations and offer more effective and reliable services. 
The statistical query under LDP mainly focuses on frequency, mean, and range.

\begin{table*}[htp!] 
\center
\caption{Comparison of frequency estimation methods}
\begin{tabular}{m{2cm}|m{2cm}|m{6cm}|m{3cm}|m{3cm}}
\hline \noalign{\smallskip}
Method & Typical papers & Description & Advantages & Disadvantages \\
\noalign{\smallskip}
\hline
\noalign{\smallskip}
\makecell[l]{Direct\\ perturbation} & \citep{kairouz2016discrete}, \citep{wang2017locally} & Perform the randomized response directly to the encoded value & Easy to perform and no additional computational cost & Poor performance when the dimension is high\\
\hline
Hash  & \citep{erlingsson2014rappor}, \citep{wang2017locally} & Hash the high dimension true value to a much smaller dimension, apply the randomized response to the hashed value & Reduce the communication cost  & Decoding process is complex and collision problem needs to be considered\\
\hline
Transformation & \citep{bassily2015local}, \citep{Apple2016} &  Transform the user value into a single bit, perform the randomized response to this bit & Small communication cost & Additional information loss during the transformation process\\
\hline
Subset selection &\citep{wang2016mutual}, \citep{ye2018optimal} & Randomly choose $k$ value to report & Good performance in the intermediate privacy region & High communication cost\\
\hline 

\label{FECompare}
\end{tabular}
\end{table*} 

\subsection{Frequency estimation}  
\begin{figure}[!htp]
\centering
\includegraphics[width=3.5in]{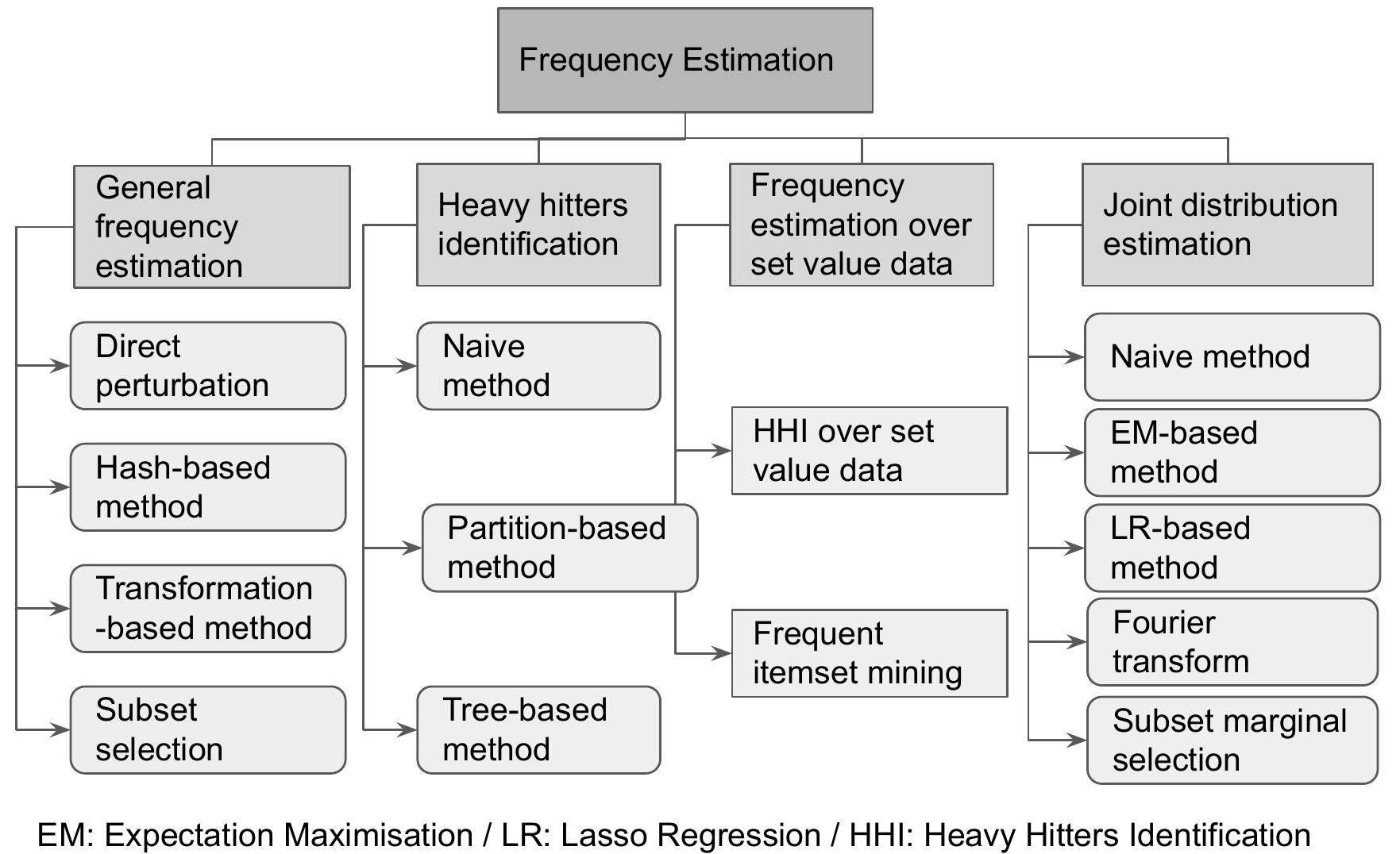}
\caption{Taxonomy tree of frequency estimation}
\label{TTFE}
\end{figure}

Frequency estimation is one of the basic statistical goals under local differential privacy protection. 
The frequency estimation mainly focuses on discrete data, such as the non-numerical data.  
We identify four types of frequency estimation: the general frequency estimation, heavy hitters identification, frequency estimation over set value data, and the joint distribution estimation. 
Fig. \ref{TTFE} shows the taxonomy tree of this section and lists the typical methods proposed in the literature for each type of frequency estimation. 
We review these methods one by one.

 
\subsubsection{General frequency estimation}
We refer the general frequency estimation to the situation that each user only has one item of an attribute. 
Specifically, 
given a set of users $U=\{u_1, \dots, u_n \}$ and an attribute $A$ with $d$ items that $A=\{I_1, \dots, I_d \}$. 
Each user $u_j$ has an item $I_i\in A$. 
The data collector aims to get the statistics of the frequency of each item in $A$, 
noted as $f_i=c(I_i)/n$, $i\in d$, where $c(I_i)$ is the number of users whose value equals to item $I_i$. 
The count number of $c(I_i)$ can be estimated by using following equation for any LDP algorithm. 
\begin{equation}
\label{Ees}
\hat{c}(I_i)=\frac{\sum_{j=1}^n \mathbf{1}_{\hat{t}_j^{\Phi}}(I_i)-nq^*}{p^*-q^*}, 
\end{equation} 
where $\hat{t}_j^{\Phi}$ is the perturbed output of user $u_j$ by an algorithm $\Phi$, 
$\mathbf{1}_{\hat{t}_j^{\Phi}}(I_i)$ is an indicator function, which is equal to $1$ when $\hat{t}_j^{\Phi}=I_i$, and $0$ otherwise. 
$p^*$ and $q^*$ are perturbation probabilities. 
 

The research focus of general frequency estimation is the perturbation mechanism design. 
Usually, the target of perturbation is to make the expectation of the estimation unbiased, and reduce the statistical variance as much as possible. This is quite challenging, especially for high dimensional data. 
Some extra measures are usually adopted to improve the accuracy. 
For example, 
the estimated item frequency can be further calibrated by incorporating the prior knowledge about the noise and item distribution \citep{jia2019calibrate}. 
Post-processing results to make them consistent by adding constraints that all estimations are non-negative and sum up to $1$ can also improve the estimation \citep{wang2019consistent, li2019estimating}. 

A pair of algorithms $<\Phi,\Psi>$, where $\Phi$ is the randomized algorithm used by each user to perturb his input value and $\Psi$ is the estimation method (e.g. Eq. \ref{Ees}) used by the aggregator, enables the aggregators to estimate the frequency of any given value $v\in R$ under local differential privacy is named \textit{frequency oracle} (FO) in the literature \citep{wang2018locally}. 
Frequency oracles are widely used as building blocks for other complicated queries and applications. 
We classify the typical FOs into following groups. \\
\\
\textbf{Direct perturbation.}
RR can be applied to the binary attribute collection directly. 
GRR is used for the attribute with $d$ possible values. 
Wang et al. \citep{wang2017locally} propose an Optimized Unary Encoding (OUE) method, 
which has a better performance when the attribute has a higher dimension.  
OUE encodes the true value as a length $d$ binary vector, $Encode(v)=[0, \dots, 0, 1, 0, \dots, 0]$, 
where only the $v$-th position is $1$. 
Each bit is performed the randomized response with probability $p=\frac{1}{2}$ and $q=\frac{1}{e^\epsilon+1}$. 

\textit{Discussion:} RR and GRR are suitable for statistics over lower dimensional data. 
OUE reduces the statistical variance for high dimensional data at the cost high communication cost. 
RR, GRR, and OUE are three most basic building blocks for other methods.
\\
\textbf{Hash-based method.} 
Erlingsson et al.'s \citep{erlingsson2014rappor} RAPPOR is a typical hash-based frequency statistical method with local differential privacy protection. 
Given a hash function $\mathcal{H}$ selected randomly from a universal hash function family $\mathbb{H}$, 
the user's value $v$ is represented as $<\mathcal{H}, \mathcal{H}(v)>$. 
RAPPOR encodes the hash value $\mathcal{H}(v)$ as a length $k$ bit vector. 
Then the randomized response is performed on each bit of the hashed vector. 
The aggregator utilizes Lasso regression \citep{tibshirani1996regression} to learn the value frequency from the collected reports to improve the estimation accuracy, which latter was proved substantially inefficient by Chai and Nayak \citep{chai2019minimax}. 
Similarly, Wang et al. \citep{wang2017locally} propose optimal local hash (OLH) method. 
OLH uses optimized choices for the range of hash functions and performs the GRR to the hashed value. 

\textit{Discussion:} Before perturbation, the hash function maps the input value with large domain size into a smaller domain size, which reduces the communication cost and statistical variance, but brings collision problem. Bloom Filter \citep{erlingsson2014rappor} and Count Mean Sketch \citep{Apple2016} are two common ways used to reduce the effect of the collision. 
To further reduce the chance of collisions, Erlingsson et al. \citep{erlingsson2014rappor} propose to assign the users permanently to $m$ cohorts, each of which has a different set of hash functions. 
All these remedies reduce the potential effect of collisions, but increases the computational complexity of the decoding process. 
\\
\textbf{Transformation-based method.}   
Given a random matrix $\Phi\in \mathbb{R}^{m\times d}$, let $\mathbf{e}_i\in \mathbb{R}^d$ be the standard basis vector with $v'th$ position equals to $1$, each column vector $\theta=\Phi \mathbf{e}_i\in \mathbb{R}^m$ of $\Phi$ represents a value $v\in A$. Each user randomly choose an index $j\in [m]$ and report $<j, \hat{\theta}_j>$ to the aggregator. 
The aggregator gets a length $m$ vector $G$ by aggregating all the reports and recovers the frequency of value $v_i$ by calculating the inner product between the aggregation $G$ and the representation of the value vector $\Phi \mathbf{e}_i$. 
Bassily and Smith \citep{bassily2015local} use a random matrix $\Phi \in \{-\frac{1}{\sqrt{m}}, \frac{1}{\sqrt{m}}\}^{m\times d}$ to map the user's value, we name it Johnson-Lindenstrauss randomized response (JLRR) in the following sections. 
Besides, Hadamard transform \citep{Apple2016, acharya2018hadamard} has been considered as well. The transformation matrix $\Phi$ is a orthogonal, symmetric $2^d\times 2^d$ matrix with $\Phi [i,j]=2^{-d/2}(-1)^{<i,j>}$. 
We name the Hadamard matrix based method as Hadamard random response (HRR).  

\begin{table*}[htp!] 
\center
\caption{Comparison among typical techniques for frequency estimation}
\begin{tabular}{cllccc}
\hline \noalign{\smallskip}
Technique & Encoding & Perturbation  & Variance & Communication \\
\hline
GRR \citep{kairouz2016discrete} & $t=v$ & $Pr[\hat{t}=v]=\begin{cases}
\frac{e^\epsilon}{e^\epsilon+d-1},& if ~t=v\\
\frac{1}{e^\epsilon+d-1},& if ~t\neq v
\end{cases}$  & $O\left(\frac{d-2+e^\epsilon}{(e^\epsilon-1)^2}\right)$ & $\log d$\\
\hline
OUE \citep{wang2017locally} & \makecell[l]{$t=[0,\cdots,1,\cdots,0]$, \\where $t[v]=1$} & $Pr[\hat{t}[i]=1]=\begin{cases}
\frac{1}{2},& if ~t[i]=1\\
\frac{1}{e^\epsilon+1},& if ~t[i]=0
\end{cases}$  & $O\left(\frac{4e^\epsilon}{(e^\epsilon-1)^2}\right)$& $d$\\
\hline
RAPPOR \citep{erlingsson2014rappor} & $\makecell[l]{r=<\mathcal{H}, t>; \\ 
\mathcal{H}\in \mathbb{H}; \\ 
t=[0,\cdots, 1, \cdots] \\ 
where~ t[i]=\begin{cases} 1, if~ \mathcal{H}(v)=1, \\ 0, ~ otherwise \end{cases} }$ & 
\makecell[l]{$Pr[\hat{t}[i]=1]= \begin{cases}
1-\frac{1}{2}f,& if ~t[i]=1\\
\frac{1}{2}f,& if ~t[i]=0
\end{cases}$,\\ where $f=\frac{2}{e^{\epsilon/2}+1}$}  & $O\left(\frac{e^{\epsilon/2}}{(e^{\epsilon/2}-1)^2}\right)$ & $\log m$\\
\hline
OLH \citep{wang2017locally} & $\makecell[l]{r=<\mathcal{H},t>;\\ \mathcal{H}\in \mathbb{H}; \\ t=\mathcal{H}(v)}$ & 
\makecell[l]{$Pr[\hat{t}=\mathcal{H}(v)]=\begin{cases}
\frac{e^\epsilon}{e^\epsilon+g-1},& if ~t=\mathcal{H}(v)\\
\frac{1}{e^\epsilon+g-1},& if ~t\neq \mathcal{H}(v)
\end{cases}$, \\ where $g=e^\epsilon +1$}  &$O\left(\frac{4e^\epsilon}{(e^\epsilon-1)^2}\right)$ & $\log n$ 
\\
\hline
JLRR \citep{bassily2015local} & $\makecell[l]{\Phi\in \{-\frac{1}{\sqrt{m}}, \frac{1}{\sqrt{m}}\}^{m\times d};\\ r=<i, t>; \\ i\in [m]; \\ t=\Phi[i, v]} $ & \makecell[l]{$\hat{t}=\begin{cases}
c_\epsilon d t,& w.p. ~\frac{e^\epsilon}{e^\epsilon+1}\\
-c_\epsilon d t,& w.p. ~\frac{1}{e^\epsilon+1}
\end{cases}$, \\where $c_\epsilon = \frac{e^\epsilon+1}{e^\epsilon -1}$}&$O\left(\frac{4e^\epsilon}{(e^\epsilon-1)^2}\right)$ & $\log m$ \\
\hline
HRR \citep{Apple2016, acharya2018hadamard}  & \makecell[l]{$ \Phi : 2^d\times 2^d$ Hadamard Matrix, \\ where $\Phi [i,j]=2^{-d/2}(-1)^{<i,j>}$;\\ $r=<i,t>$; \\ $i\in [2^d]$; \\ $t=\Phi [i, v]$   } & $Pr[\hat{t}=1]=\begin{cases}
\frac{e^\epsilon}{e^\epsilon+1},& if ~t=1\\
\frac{1}{e^\epsilon+1},& if ~t=-1
\end{cases}$  & $O\left(\frac{4e^\epsilon}{(e^\epsilon-1)^2}\right)$ & $O(1)$  \\
\hline
\label{BuildingBlocks}
\end{tabular}
\end{table*}


\textit{Discussion:} Transformation-based method transforms the user's value from $d$ bits to only $1$ bit.  On the one hand, it reduces communication cost significantly. On the other hand, one bit data might not represent the complete input information. The accuracy can be affected, especially when the privacy budget $\epsilon$ is big, the information loss during the transformation process dominates the statistic error.   
\\
\textbf{Subset selection}
The general idea of subset selection is randomly select $k$ items within the domain size. 
The user's true item is included in the $k$-subset with a specified probability. 
Wang et al. \citep{wang2016mutual} get the optimal subset size $k=\lfloor \frac{d}{e^\epsilon +1}\rfloor$ or $k=\lceil \frac{d}{e^\epsilon+1 }\rceil$ by minimizing variance. 
Later, they extend the $k$-subset mechanism to tackle discrete quantitative data with arbitrary distance metric by proposing a variant of the Exponential Mechanism (EM) using a trivial output as padding \citep{wang2019local}.
Ye and Alexander \citep{ye2018optimal} provide a much detailed analysis of the $k$-subset mechanism in the medium privacy regime and provided a tight lower bound on the minimax risk. 


\textit{Discussion. } Subset selection method shows a good performance in the intermediate privacy region (e.g. $\log 2 \leq \epsilon \leq \log (d-1)$) \citep{wang2016mutual} compared with GRR. 
But the communication cost is relatively high, especially when $\epsilon$ is small. \\ 
\\ 
\textbf{Summary}
The frequency estimation is the most basic statistic and the most mature study over local differential privacy. 
The main challenges of frequency estimation are to alleviate the effect of high dimensional data and reduce  communication cost without sacrificing statistical accuracy too much.   
To provide a much intuitive comparison among these techniques, we summarise these methods in Table \ref{FECompare} and list the typical techniques in Table \ref{BuildingBlocks}. 
The perturbation methods are adaptively selected according to different application requirements. 
For example, GRR is preferred for low dimensional data statistic and transformation-based methods are more used in situations requiring small communication, such as in the IoT area.

\subsubsection{Heavy hitters identification} 
The aim of heavy hitter identification is to find the items with frequency over a threshold. 
For example, find the items whose frequency is larger than $10\%$. 
Heavy hitter identification is a basic research problem in data analytics with many applications, such as frequent item mining, trend monitoring, and marketing analysis.
Heavy hitters can be identified by using frequency oracle directly, we name it naive method. 
Or, we combine other algorithms with frequency oracle to make the process much efficient, such as partition-based method and tree-based method. 
Table \ref{heCompare} summaries these methods.\\
\\
\begin{table*}[htp!] 
\center
\caption{Comparison of heavy hitter identification methods}
\begin{tabular}{p{2cm}|p{2cm}|p{6cm}|p{3cm}|p{3cm}}
\hline \noalign{\smallskip}
Method & Typical papers & Description & Advantages & Disadvantages \\
\noalign{\smallskip}
\hline
\noalign{\smallskip}
Naive method & \citep{kairouz2016discrete}, \citep{wang2017locally} & Apply the frequency oracle, compute frequency of all times, then find the top-$k$ frequent ones & Easy to perform and no additional computation & Not efficient\\
\hline
Partition-based method & \citep{fanti2016building},\citep{wang2019locally} & Partition the true vector into small non-overlapping or overlapping segments, randomly choose one or few segments, apply randomized response to each bit of the vector & High efficient for heavy hitter estimation & Additional computation cost, low accuracy \\
\hline
Tree-based method  & \citep{bassily2017practical}, \citep{wang2018privtrie}  & Iteratively statistic the vector, prune the less frequent prefixes before growing the tree. & Find the frequent items efficiently, donot need to know the domain size & Need multiple iteration  \\
\hline
\label{heCompare}
\end{tabular}
\end{table*} 
\\
\textbf{Naive method.} 
As the heavy hitters are the items with the frequency over a threshold, 
the aggregator can collect the user's data and compute the frequency for each item using frequency oracle, then find out the heavy hitters. 

\textit{Discussion.} Naive method is an effective solution for heavy hitter identification for low dimensional data. But it is inefficient for high dimensional data. For example, assume there are $1000$ websites in total, the aggregator wants to know the top-$10$ frequently visited ones, he needs to query $2^{1000}$ times to get the answer just using frequency oracle. 
And the statistical variance would be very high, which reduces the statistical accuracy significantly.\\
\textbf{Partition-based method}
Partition-based method tries to find the frequent ones without going through all the possible values. Specifically, the user's data is encoded as a binary vector using one-hot encoding.  
Each user reports a segment of the perturbed vector to the aggregator. 
If the user's vector is partitioned into $g$ segments, each user only needs to report $s=d/g$ bits. 
The principle is that if a value is frequent, the segment of the value is also frequent. 
The aggregator finds the frequent strings in each segment denotes as $C_i$ and then calculates the Cartesian product of $C_i$ as $C=C_1 \times C_2\times ...\times C_g$. The frequent items are found in the candidate set $C$. 
To further reduce the size of the candidate set $C$, Fanti et al. \citep{fanti2016building} let the users report two segments randomly instead of one. 
A potential problem of partition is that if the number of segment $g$ is big, there will be a small group of users report the same segment, which reduces the statistical accuracy. Wang et al. \citep{wang2019locally} solve this problem by having the segments overlapping. That is, they let the users in each group report a prefix of their value with predefined length. Then the aggregator finds the frequent items iteratively. 
Kim et al. \citep{kim2018learning} propose a similar method to find frequent words from users' keystroke data. Specifically, each user appends a hash value to the word (enable integrity checking) and sends one random segment (the segment can start form any position of the word string, and there are partial overlapping between segments) to the server. 
The server selects the frequent strings of the segment and combines candidate segments per position side-by-
side following the chain rule. 

\textit{Discussion.} Partition-based method increases the efficiency of heavy hitters estimation by reducing the query times from $2^d$ to $2^sg$ (if report one segment). 
However, it increases the other computational cost, such as the construction of the candidate set. 
Besides, the candidates obtained from the segments often do not correspond to real terms, which affects the accuracy of heavy hitter estimation. \\
\textbf{Tree-based method}
The tree-based method is mainly used for 'string data', such as trajectory and English word. 
The frequent item is identified by iteratively constructing a tree under LDP, each node represents a prefix of an item and less frequent prefixes are pruned during each iteration.
Similar to partition-based method, the principle of tree-based method is that all prefix of a frequent item must be frequent as well, which enables effective pruning. Therefore, the statistical computation of heavy hitters is much more efficient. As each item includes multiple characters, the domain size is very high,  
the estimated frequency may not be as accurate as expected. To solve this problem, Bassily et al. \citep{bassily2017practical} invoke the local randomizer twice in the full protocol, once during the pruning process where the high-frequency items are identified, and a second time during the estimation phase, invokes the frequency oracle once more on those particular items to enable the protocol to get a better estimation. 
Wang et al. \citep{wang2018privtrie} present a candidate set construction method for each node on the tree, which restricts that each user can only report one time to the nodes on his own path to save the privacy budget. 

\textit{Discussion:} Tree-based method can find the frequent items efficiently and users do not need to know the domain size of data. The drawback is the algorithm needs multiple iterations, which increase the communication cost and delay. Besides, currently, the main method to reduce the statistical variance is to partition users to disjoint groups. When the data dimension is large, the insufficient number of users in each group reduce the accuracy as well. \\
\\
\textbf{Summary.} The main challenge for heavy hitter identification is how to find the heavy hitters efficiently, and with accuracy. The basic idea of current method is to remove some infrequent items step by step. On the one hand, it solved the efficiency problem, but on the other hand, it introduces new problems, such as complex computation or high communication costs. 
To further improve the performance of heavy hitter identification is remain a challenge. 

\subsubsection{Frequency estimation over set value data}

\begin{table*}[htpb]  \centering
  \caption{Typical papers for set-value data collection}\label{tab-parameter}
  \begin{tabular}{m{2cm}|m{2cm}|m{4cm}|m{4cm}|m{4cm}}
\hline
    Protocol & Task & Key methods & Advantages & Disadvantages    \\
\hline
LDPMiner \citep{qin2016heavy} & Find frequent items & Refine the candidate set & Easy to implement & Only works when $k$ is small \\
\hline
PrivSet \citep{wang2018privset} & Find frequent items& Output a subset of the items & The result can be quite accurate & More communication cost and inefficient to chose the size of subset \\
\hline
SVSM \citep{wang2018locally}  & Find frequent itemsets & Construct candidate sets using guessed frequency based on the estimated frequent items & Small size of the candidate sets reduces the statical variance & Extra error introduced due to the partitioning of the users and guessed frequency\\
\hline
\end{tabular} \label{FIMC}
\end{table*}
Set-value data contains a collection of items.
Specifically, each user has a set of up to $\ell$ items (e.g., web pages browsed, movies watched, locations visited) instead of one. 
Two types of statistical queries are considered in the literature, find the items with frequency over $f$ and find the itemsets with frequency over $f$, 
which correspond to two data mining tasks, heavy hitters identification and frequent itemset mining (FIM). 

\textit{Heavy hitters identification over set value data.} The main challenge for identifying the heavy hitters over set value data is that the set-value data has heterogeneous size. 
That is, each user may have $0$ to $l$ items, which makes it difficult to access the item sample probability, then make accurate frequency estimation difficult. 

Currently, the problem of heterogeneous size is solved by trimming and padding \citep{qin2016heavy, wang2018privset}. Specifically, the aggregator assumes the variable $m$ to be the largest number of user's size of the set-valued data. If the number of items in user's set-valued data is beyond $m$, the data is simply truncated to $m$ items. If the number of items in user's set-valued data is under $m$, dummy items are added to achieve $m$. Then each user samples one item randomly to report. 
To enhance accuracy, 
Qin et al. \citep{qin2016heavy} propose LDPMiner, which is a two-phase mechanism. 
A candidate set of top-$k$ frequent items is identified using a portion of the privacy budget in the first phase. The remaining budget is used to refine the candidate set in a second phase. 
Wang et al. \citep{wang2018locally} utilize the privacy amplification of sampling to save privacy budget. 
As the privacy mechanisms used in both \citep{qin2016heavy,wang2018locally} are the same as the LDP mechanism for general frequency estimation, 
the privacy budget is unable to be fully utilized for set-value data collection. 
Wang et al. \citep{wang2018privset} propose to output a subset of the items using Exponential Mechanism without splitting the privacy budget, which results in a better accuracy, but how to choose the size of subset efficiently is still an open issue. 

\textit{Frequent itemset mining.} Frequent itemset mining aims to find the itemsets with a frequency over a given threshold. These itemsets include rich information, which help the service provider improve the quality of services, such as understanding the users' usage pattern and predicting the user behaviour \citep{adar2007we}. 
While the main challenge for frequent itemset mining is that it has an extremely large domain, which results in big statistical variance. Assume there are totally $d$ different items, the number of possible values of each user's set-valued data is $\binom{l}{m}+\binom{l}{m-1}+\dots+\binom{l}{0}$. 
Therefore, reducing the candidate set size is a major research issue for frequent itemset mining. 
While another challenge is how to encode the input under the local DP setting. 

Due to the challenges of FIM, only Wang et al. \citep{wang2018locally} propose a solution under LDP setting. 
Specifically, they construct the frequent itemset candidate set based on the identified frequent items. 
And then reduce the size of the candidate set by selecting the itemsets with the highest guessing frequencies.  
The intuition is that it is very unlikely that a frequent itemset is composed of several infrequent items. 
However, the splitting of the privacy budget and guessed frequency increase the statistical errors, which needs to be further reduced.  \\
\\
\textbf{Summary.} The key feature of set-valued data is that the user's data size are heterogeneous. 
Users whose data size under $m$ add items to achieve the padding size. 
As the added items are fake ones, which cannot contribute to the statistic. 
Therefore, the effective samples are reduced with the increase of the padding size $m$. 
However, if the $m$ is too small, the sampled item may not represent the user's real data. 
The choose of padding size $m$ in the literature is subjective without theoretical support, which is still an open problem. 
The key problem of collecting the set-value data is that it has a very large domain, especially for frequent itemset mining. 
The main challenge is how to reduce the size of candidate set. 
Table \ref{FIMC} summaries the typical protocols for set-value data collection.

\subsubsection{Joint distribution estimation}

\begin{table*}[htp!] 
\center
\begin{threeparttable}[b]
\caption{Comparison of methods for joint distribution estimation}
\begin{tabular}{l|l|lllcll}
\hline \noalign{\smallskip}
Method & \tabincell{l}{Typical \\paper }& \tabincell{l}{Computation \\complexity}& \tabincell{l}{Communication \\cost}& Variance & Advantages & Limitations   \\
\hline
 Naive method & --& High & High & $ 2^d\cdot Var\tnote{1}$ & $\bullet$ Compute any $k$-way marginals & \tabincell{l}{$\bullet$ High variance \\$\bullet$ Inefficient}\\
\hline
 EM & \citep{fanti2016building},\citep{ren2018textsf} & High & High & $ 2^d\cdot Var$ &  \tabincell{l}{$\bullet$ Compute any $k$-way marginals \\$\bullet$ Much accurate } & \tabincell{l}{$\bullet$ High variance \\$\bullet$ Inefficient}\\
\hline
\tabincell{l}{Lasso \\regression} & \citep{ren2018textsf} & Medium & High & $ 2^d\cdot Var$& \tabincell{l}{$\bullet$ Compute any $k$-way marginals \\$\bullet$ Much efficient} & $\bullet$ High variance \\
\hline
\tabincell{l}{Fourier\\ Transformation} & \citep{cormode2018marginal} & Medium&Low & $\sum_{i=1}^t \binom{d}{i}\cdot Var $ &$\bullet$ Low communication cost  & $\bullet$ Predefine $k$ \\
\hline
\tabincell{l}{Subset\\ selection} & \citep{zhang2018calm} &  Medium &Medium & $\frac{m}{n}\cdot 2^\ell\cdot Var$ & \tabincell{l}{$\bullet$ Compute any $k$-way marginals \\ $\bullet$ Low variance} & $\bullet$ Extra errors \\
\hline

\label{CMJDE}
\end{tabular}
\begin{tablenotes}
  \item[1] $Var$ is the variance of a single cell in the full contingency table
\end{tablenotes}
\end{threeparttable}
\end{table*} 
Apart from the frequency estimation for a single attribute, 
a more natural and general setting is that each user has multiple attributes, and the aggregator is interested in the joint distribution of some of these attributes, which names marginals. 
For example, the aggregator might want to study the association between displayed advertisements and recently-installed software applications. 
A variety of fundamental inference and machine learning tasks also rely on accurate marginals capturing the correlations \citep{ye2018optimal,nie2018classification,yilmaz2019locally}. 
We summarise three methods for joint distribution estimation as follows. \\
\\
\textbf{Naive method}
The simplest method is that take the user's record with multiple attributes as a single item. 
Then get the joint distribution of all attributes by performing frequency oracles. 
$k$-way marginal table can be computed accordingly.  
However, directly calculate the whole marginal table has two obvious shortcomings. 
\begin{itemize}
\item The statistic variance is very high, which is proportional to the domain size of the statistic. 
\item All the value frequency of attributes need to be calculated, which cause a high computation cost. The time complexity and space complexity grow exponentially with the number of attributes $d$ and can be prohibitively expensive.
\end{itemize}
\textbf{Expectation maximisation (EM)-based method.}
EM method allows the users to report the value for each attribute separately with split privacy budget, then the marginal table is reconstructed based on the collected noisy value.   
Specifically, EM initializes the probability density distribution of the joint attributes $p(a_1, a_2, ..., a_d)$ as a uniform distribution. 
And then, update it iteratively by calculating $p_{\gamma+1}(a_1, a_2, ..., a_d)=\frac{1}{N}\sum_{i=1}^Np_{\gamma}(a_1, a_2, ..., a_d|\hat{\mathbf{t}}_i)$, 
where $p_{\gamma}(a_1, a_2, ..., a_d|\hat{\mathbf{t}}_i)$ is the posterior probability, which can be computed by the Bayes's Theorem, $\gamma$ is the number of iteration, and $\hat{\mathbf{t}}$ is the perturbed report. 
Let $A_j$ be the joint attribute $a_1, a_2, ..., a_d$, then,
\begin{equation}
p_{\gamma}(A_j|\hat{\mathbf{t}}_i)=\frac{p_{\gamma}(A_j)p(\hat{\mathbf{t}}_i|A_j)}{\sum_{A_j \in Domain} p_{\gamma}(A_j)p(\hat{\mathbf{t}}_i|A_j)}
\end{equation}
The iterative process stops when the maximum difference between two estimations is smaller than the specified threshold. 
This method is first introduced by Fanti et al. \citep{fanti2016building} for estimating joint distribution for two attributes, and then generalized by Ren et al. \citep{ren2018textsf} to handle multiple attributes. 

\textit{Discussion:} Similarly as the naive solution, EM method has the advantage of being able to compute $k$-way marginals for any $k$. 
But the privacy budget is split into each attribute, which results in a large variance. 
Therefore, EM algorithm is not suitable for the high dimensional data, as the domain size of the joint attributes increases exponentially with the increase of dimension $d$ as well. 
EM algorithm achieves a better accuracy than the naive method, however, does not consider the problem of time and space complexity either. 
Besides, the EM algorithm relies on the initial value (initialized distribution), which determines the speed of convergence. Both work \citep{fanti2016building} and \citep{ren2018textsf} initialize the joint distribution as uniform distribution, which is obviously not optimal for convergence. \\
\textbf{Lasso regression-based method}
Ren et al. \citep{ren2018textsf} model the joint distribution estimation problem as linear regression. 
The method estimates the number of each bit according to the general randomized response theory as the response variable $y$. The distribution is estimated from the noised sample space by taking advantage of linear regression $y=\mathbf{M}\beta$, where $\mathbf{M}$ is the predictor variables and represented as Bloom filters. 
For the $d$-dimensional data, let $m_i$ be the length of Bloom Filter of each attribute, the size of $\mathbf{M}$ is $\prod_{i=1}^d m_i$, which includes all the possible combinations. 
That is $\mathbf{M}=[\mathcal{H}_1(a_1)\times \mathcal{H}_2(a_2)\times ...\times \mathcal{H}_d(a_d)]$. 
The joint distribution estimation can be derived by reshaping the coefficient vector $\beta$ into the $d$-dimensional matrix. 

\textit{Discussion.}  Lasso regression-based method improves the efficiency of the estimation, 
but does not have any contribution to the estimation accuracy. Based on the estimated distribution and correlations, the extensions of generating synthetic dataset are proposed in the literature \citep{ren2018textsf, yang2017copula, wangt2019locally}. \\
\textbf{Fourier Transform}
Cormode et al. \citep{cormode2018marginal} apply Hadamard Transformation to publish $k$-way marginals in the local setting. 
Given a vector $t\in \mathcal{R}^{2^d}$ and a Hadamard metric $\Phi$, the vector $t$ can be represented by the $2^d$ Hadamard coefficients in the vector $\theta=\Phi t$. 
Assume the vector $t$ has the single $1$ at index $\ell$, the user only need to randomly select a coefficient $\theta_i=\Phi_{i,\ell}$ to report by randomized response. 
And the aggregator only needs $\sum_{j=0}^k\binom{d}{j}$ coefficients to calculate $k$-way marginal. 

\textit{Discussion.} The advantages of Fourier transform are that it saves communication cost and has much lower variance when $k$ is small. However, when $k$ is large, a large number of coefficients ($O(d^k)$) need to be estimated. Besides, the method only suitable for binary data and the number of $k$ needs to be predefined. \\ 
\textbf{Subset Marginal Selection}
Instead of reporting the value of all the attributes, 
an alternative choice is that let each user report a subset of the attributes. 
For example, Zhang et al. \citep{zhang2018calm} propose CALM, which borrows the idea of PriView method \citep{qardaji2014priview} for publishing marginal under the centralized DP setting. 
CALM publishes a synopsis of the dataset. 
The synopsis takes the form of $m$ size-$\ell$ marginals, which are called views. 
The aggregator chooses a set of $m$ marginals and FO protocol to be used, then assigns each user to one of the marginals. 
The user only needs to report the values of the attributes included in the marginal he needs to report. 
The aggregator calculates the marginal tables according to users' report and constructs the $k$-way marginal.

\textit{Discussion:} The key highlights of the method are that it can release any-way marginal table without calculating the full marginal, and it can deal with the non-binary attributes. 
However, it introduces too much error. 
Besides the noise error caused by the privacy-preserving, 
each estimation of the value is only based on the part of the users, which introduces the sampling error. 
Also the marginal construction process causes the construction error. \\
\\
\textbf{Summary.} Joint distribution query has a much higher statistical variance than the frequency estimation for the single attribute. 
The marginal statistics (statistic the distribution over any number of attributes in the domain) make it much more complicated. 
The main research question is how to reduce the computation and communication cost while enabling any-way marginal statistics. 
To make the current research clear, we show the comparison of the existing methods in Table. \ref{CMJDE}.

\subsection{Mean value estimation}

 \begin{table*}[htp!] 
\center
\caption{Comparison of single numerical attribute collection}
\begin{tabular}{clcll}
\hline \noalign{\smallskip}
Method & Perturbation & Variance & Advantage & Disadvantage \\
\hline

Duchi's \citep{duchi2018minimax}
 & Extreme value 
 

& $ (\frac{e^{\epsilon}+1}{e^{\epsilon}-1})^2 $   & \tabincell{l}{ Easy to encode  Accurate statistics \\for low  privacy regime} & \tabincell{l}{ The variance cannot be reduced \\with the increase of the $\epsilon$} \\
\hline

PM \citep{wang2019collecting}
 &  Distribution 
 



&$\frac{4e^{\epsilon/2}}{3(e^{\epsilon/2}-1)^2}$ & \tabincell{l}{ Good performance for the whole \\range of privacy regime} & \tabincell{l}{ Sophisticated computation and \\hard to encode}\\
\hline
\label{MVE}
\end{tabular}
\end{table*}

 \begin{table*}[htp!] 
\center
\caption{Comparison of multiple numerical attributes collection}
\begin{tabular}{clcc}
\hline \noalign{\smallskip}
Method & Perturbation & Communication & Error \\
\hline

Duchi's \citep{duchi2018minimax}
 &  Extreme value for high dimensional data

& $O(d)$ & $O( \frac{  \sqrt{d \log d } }{\epsilon \sqrt{n}} )$ \\
  
\hline

Harmony \citep{nguyen2016collecting} & 

\tabincell{l}{ 

Randomly sample one attribute $A_i$ to report.  \\
Extreme value perturbation.}

  & $O(1)$ & $O( \frac{  \sqrt{d \log (d/\beta } }{\epsilon \sqrt{n}} )$\\
  \hline
PM \citep{wang2019collecting} & \tabincell{l}{ Randomly sample $k$ attributes \\ Distribution perturbation} &$O(k)$ & $O( \frac{  \sqrt{d \log (d/\beta } }{\epsilon \sqrt{n}} )$ \\

\hline
\label{MVEM}
\end{tabular}
\end{table*}
Different from the categorical attributes, the domain of the numeric attribute is a range, and there are an infinite number of values. 
Therefore, the general methods used for the categorical attribute are not suitable for numeric attributes. 
The main application of numerical data collection is mean value estimation \citep{ding2017collecting}. 
Assume the input domain of the numerical values is $\mathcal{D}=[a,b]$, 
each user hold a value $v_i\in \mathcal{D}$, the aggregator collects user's report $\hat{t}_i$ and simply compute the average $\frac{1}{n}\sum_{i=1}^n \hat{t}_i$. 
The research focus is on how to get a much accurate mean value with a much smaller statistical variance. 
\subsubsection{General mean value estimation} 
We refer the general mean value estimation to the situation that each user has a single numerical value for each attribute. 
Currently, there are mainly two types of randomization for general mean value estimation, 
extreme values perturbation and distribution perturbation. \\
\\
\textbf{Extreme values perturbation.} The main idea of extreme values perturbation is that each user reports one of two extreme values with a specified probability depending on the input value $v$, no matter what $v$ is. 
The probability ensures the expectation of the estimated value $\hat{v}$ equals to true value $v$, which provides unbiased mean value estimation. Given an input $t\in [-1,1]$, Duchi et al. \citep{duchi2013local,duchi2014privacy,duchi2018minimax} makes the output equals to either $\frac{e^\epsilon+1}{e^\epsilon-1}$ or $-\frac{e^\epsilon+1}{e^\epsilon-1}$, with probability $\frac{e^\epsilon-1}{2e^\epsilon+2}\cdot t +\frac{1}{2}$ and $-\frac{e^\epsilon-1}{2e^\epsilon+2}\cdot t +\frac{1}{2}$ respectively. 
Besides, Duchi et al. proposed solution for multiple numerical attribute collection,
which takes as input a tuple $\mathbf{t} \in [-1, 1]^d$ of a user,
and outputs a perturbed vector $\hat{\mathbf{t}} \in \{-B, B\}^d$, 
where $B$ is a constant decided by $d$ and $\epsilon$. 

\textit{Extensions.} For multiple attributes statistics, based on Duchi et al. solution, Nguyen et al. \citep{nguyen2016collecting} propose Harmony, which randomly chooses one attribute to report instead of reporting all attribute values. Harmony reaches a similar statistical variance with Duchi's but with much smaller communication cost. 
Later Akter and Hashem \citep{akter2017computing} consider that the user may have a different privacy concern, and extend Duchi et al.'s solution to satisfy personalized privacy setting. 
Wang et al. \citep{wangt2019locally} adjust the output probability of Duchi et al.'s method for multiple attributes collection, making it satisfy $(\epsilon, \delta)-LDP$, which gets a smaller variance.  

\textit{Discussion.} Duchi et al.'s solution offers considerable smaller variance when $\epsilon$ is small, 
but not very well when $\epsilon$ becomes large, because it's worst-case variance $O((\frac{e^\epsilon+1}{e^\epsilon-1})^2)$ is always over $1$ no matter how big the privacy budget is. 
Kairouz~\emph{et~al.}~\citep{kairouz2014extremal} also prove that this binary mechanism (two outputs) is not always optimal as $\epsilon$ increases. Therefore, more possible outputs might be explored in the future.\\
\textbf{Distribution perturbation.} The idea of distribution perturbation is that model the output of the value as a continuous distribution. Specifically, define an output domain $[-s, s]$, which is broader than the input domain $[-1, 1]$. 
For each value $v$, there is an associated rang $[\ell (v), r(v)]$, where $-s\leq \ell (v) < r(v) \leq s$. 
The user with value $v$ outputs the value $\hat{v} \in [\ell (v), r(v)]$ with higher probability, and output other value with lower probability. 
The method was first proposed by Wang et al. \citep{wang2019collecting}, named Piecewise Mechanism (PM), 
to estimate the mean value, 
which achieves a much smaller variance compared with Duchi et al.'s solution, especially when $\epsilon$ is big. Extending the single numerical value into multi-dimensional numeric attributes is also considered in the paper \citep{wang2019collecting}. The general idea is that randomly sampling $k$ dimensional data to report instead of all the values.

\textit{Extensions.} Li et al. \citep{li2019estimating} propose a similar method with PM, named Square Wave mechanism, to reconstruct the distribution instead of calculating the mean value.  
Given a input domain $[0, 1]$ and output domain $[-b, 1+b]$, the user output $\hat{v} \in [v-b, v+b]$ with probability $\frac{e^\epsilon}{2be^\epsilon+1}$ and output other values with probability $\frac{1}{2be^\epsilon+1}$. Parameter $b$ is chosen by maximizing the upper bound of mutual information between the input and output of the reporting. As input values are always at the center of high probability region, Square Wave mechanism cannot provide unbiased estimation for mean value.

\textit{Discussion.} The distribution perturbation method gets a much better accuracy for the high privacy regime. The short-comings are that it brings sophisticated computation and hard to encode due to the unlimited number of possible output.\\
\\    
\textbf{Summary.} Collecting numeric data has not been addressed sufficiently under local differential privacy protection. 
Currently, 
Duchi's method and PM are two 
main solutions. Other extensions based on these two mechanisms are all target on mean value estimation as well. Table \ref{MVE} and Table \ref{MVEM} show the typical methods for single numerical value collection and multiple numerical data collection respectively. 
Due to the big bias for the single data record, the designation of algorithms for statistics, such as max, min, and quantile, is quite challenging. 
\subsubsection{Mean value estimation over key value data}
Key-valued data is a popular NoSQL data model, and is widely encountered in practice.
It is a hybrid data type that contains both categorical attributes and numerical attributes.  
And each user may have multiple key-value pairs. 
Table \ref{MovieData} shows an example of key-value data. 
The key is the movie name, which also can be represented as the movie ID. 
The value is the movie rating. 
Both movie name and rating should be perturbed to prevent the attacker from knowing which movie the user watched or rated and how many scores the user rated the specific movie. 

\begin{table}[htp!] 
\center
\linespread{1.3}\selectfont
\caption{Movie dataset}
\begin{tabular}{|c|l|c|}
\hline
User& Movie & Rating \\
\hline
 \multirow{3}*{$u_1$} & The godfather & $9.2$ \\
 \cline{2-3} & Cats & $2.8$ \\
 \cline{2-3} & Knives Out & $8.1$\\
 \hline
 \multirow{5}*{$u_2$} & Little Women & $8.3$ \\
 \cline{2-3} & Cats & $8.8$ \\
 \cline{2-3} & Uncut Gems & $8.1$ \\
 \cline{2-3} & Star Wars & $6.9$ \\
 \cline{2-3} & Joker & $8.7$ \\
 \hline
\label{MovieData}
\end{tabular}
\end{table} 

Usually, we query the frequency of keys and mean of values under each key simultaneously. 
The frequency estimation of keys can be achieved by applying the frequency oracle directly. 
The main challenge is that there is an inherent correlation between key and values, which if not harnessed, will lead to poor utility when querying the mean values. 
To keep the correlation between the key and value, the solution that perturbs the value based on the perturbation result of the key is proposed. 
However, if a user reports a key that does not exist in his local data, a fake value has to be generated to guarantee the indistinguishability.  

\begin{table*}[htpb]  \centering
  \caption{Comparison of Key-value data statistical methods}\label{tab-parameter}
  \begin{tabular}{cllll}
\hline
Method & Typical paper & Key points & Advantages & Disadvantages   \\
\hline
\multirow{2}*{\tabincell{l}{Perturb the value \\based on the \\perturbation result \\of the key}}& PrivKVM \citep{DBLP:conf/sp/YeHMZ19} & \tabincell{l}{$\bullet$ Generate fake value \\randomly from $[-1, 1]$ \\$\bullet$ Iteratively update fake value \\as the estimated mean value}    & $\bullet$ Unbiased mean value estimation & $\bullet$ Multiple rounds iteration \\
\cline{2-5}& PCKV \citep{gu2019pckv} & \tabincell{l}{$\bullet$ Generate fake value as either\\ $1$ or $-1$ \\$\bullet$ Utilize tight budget \\composition and optimized \\budget allocation }  & \tabincell{l}{$\bullet$ Fake values achieves expected \\zero summation without iteration\\ $\bullet$  Less privacy budget consumption} & \tabincell{l}{$\bullet$ Big variance for a smaller \\population} \\
\hline
\end{tabular} \label{KVDC}
\end{table*}

Ye et al. \citep{DBLP:conf/sp/YeHMZ19} propose to generate the fake value randomly from $[-1,1]$ at beginning, and then iteratively update the fake value to improve the estimation of the mean value. Gu et al. \citep{gu2019pckv} generate the fake
values as $-1$ or $1$ with probability $0.5$ respectively, the expectation of which is zero and has no contribution to the value summation statistically. Besides, they proposed an optimized privacy budget allocation scheme to improve the utility further. 
Sun et al. \citep{sun2019conditional} propose a new idea in their preprint article. 
They encode the continuous numerical value as two extreme values $-1$ and $1$, 
then randomly sample the value from the perturbed space, which constructed by all the possible combinations between the key and the value.   
To make full use of the original pairs, Gu et al. \citep{gu2019pckv} propose to use padding and sampling protocol \citep{wang2018locally} instead of sampling the pair directly from the domain. \\
\\
\textbf{Summary.} There are two new problems need to be solved in key-value data statistic. 
One is how to keep the correlation between the key and the value. The second problem is how to generate fake value with a minimal impact on the final statistical results. 
Table \ref{KVDC} illustrates the advantages and disadvantages of current solutions.

\subsection{Range query estimation}

A range query counts the fraction of a population having the value within a specified interval. 
For example, 'What percentage of people are between the ages of $30$ and $40$?'. 
The query can be estimated directly via point estimates. 
That is the aggregator simply sums up estimated frequencies for every item ($30,31,...,40$) in the range. 
This works tolerably well for short ranges over small domains, but rapidly degenerates for larger inputs due to the accumulated variance. 
As the variance grows linearly with the interval size. 
Assume the statistical variance of each point is $Var$, the variance of a range with length $m$ is $m\cdot Var$. Currently, the main method to solve this problem is hierarchy-based method, which bounds the variance by a polylogarithm of the length of the range.  \\
\\
\textbf{Hierarchy-interval method.} 
The general idea of hierarchy-interval based method is to construct a b-ary tree of height $h$. 
Each node corresponds to an interval, and has $b$ children, which corresponds to $b$ equally sized subintervals. The users are randomly partitioned into $h$ groups, the users in group $G_i$ report to the nodes in level $i$. The aggregator estimates the fraction of the input at each node and answers the range queries by aggregating the nodes from the decomposition of the range. 
Hierarchy-interval based method rescales the error logarithmically with the length of the range. 
That is, compared with naive method, it reduces the variance from $O(m)\cdot Var$ to $O(\log_b m)\cdot Var$. 

Cormode et al. \citep{kulkarni2019answering,cormode2019answering} first discuss the range query under local differential privacy setting.  
They utilized B-adic intervals to decompose the data domain and further improved the accuracy by post-processing the result that makes the number of child nodes sum to the number of the parent node. 
Wang et al. \citep{wang2019answering} discuss the problem of multi-dimensional analytical queries. 
For example, \\
\textbf{SELECT}  COUNT(*) \\
\textbf{FROM} D \\
\textbf{WHERE} Age $\in [30, 40]$ AND Salary $\in [50K, 150K]$\\
They applied the Hierarchical-interval method to each attribute. 
Therefore, the worst-case squared error for $d$ ordinal dimensions is reduced from $O(m^d )$ to $O(\log^d m)$. 
Based on the proposed algorithm in \citep{wang2019answering} for answering multi-dimensional analytical queries, they proposed and demonstrated a middleware solution DPSAaS, which provides differentially private data-sharing-and-analytics functionality as cloud services \citep{xu2019dpsaas}. 
\\
\textbf{Hierarchy-coefficient method.} Haar wavelet transforms (HWT) \citep{stollnitz1996wavelets}, a popular technique for processing one-dimensional ordinal data, has been used to summarize data for the purpose of answering range queries \citep{xiao2010differential}. Specifically, given a vector $\mathbf{t}$ with length $m$, HWT wavelet constructs a binary tree with $\log m$ levels. 
The leaves are corresponding to the entries in $\mathbf{t}$. 
It then generates a wavelet coefficient $c$ for each internal node, such that $c=\frac{c_l-c_r}{2}$, 
where $c_l (c_r)$ is the average value of the leaves in the left (right) subtree. 
An additional coefficient $c_0$ (referred to as the base coefficient) is produced by taking the mean of all leaves. 
Given the Haar wavelet coefficients, any entry $v$ in $\mathbf{t}$ can be easily reconstructed by following equation. 
\begin{equation}
v=c_0+\sum_{i=1}^l(g_i\cdot c_i),
\end{equation}
where $c_i$ is the ancestor of $v$ at level $i$, and $g_i$ equals $1 (-1)$ if $v$ is in the left (right) subtree of $c_i$. 
Similar to Hierarchy-interval method, Cormode et al. \citep{cormode2019answering} let each user choose one level to report and use HRR to perturb the coefficient. 
The noisy coefficient is collected to estimate the count of the item. 
The range query can be answered by simply summing the count of all items in the range, which needs at most $2h$ coefficients. Therefore, the variance of the query result is also bounded by a polylogarithm of $m$.\\
\\
\textbf{Summary.} Range query is a very common database operation that retrieves the recodes where some value within an interval. 
Simply aggregate the point estimate results in big statistical variance, hierarchy-based method is the only way to solve this problem under LDP protection in the literature. 
More structure of data storage that might contribute to the range estimation needs to be explored.

\begin{table*}[htpb]  \centering
  \caption{Comparision of supervised learning methods}\label{tab-parameter}
  \begin{tabular}{m{2cm}|m{4cm}|m{4cm}|m{5cm}}
\hline
    Typical papers & Method & Advantages & Disadvantages    \\
\hline
 \citep{yilmaz2019locally} & Bind two attributes as one, then apply the general FO & Support both numerical and categorical attributes & High statistical variance\\
\hline
 \citep{xue2019joint} & Perturb the value sequentially using subset selection method & Much accurate statistics & High communication cost and only support categorical attributes\\
\hline
\end{tabular} \label{NBBCC}
\end{table*}

\begin{table*}[htpb]  \centering
  \caption{Comparision of unsupervised learning methods}\label{tab-parameter}
  \begin{tabular}{m{2cm}|m{4cm}|m{1cm}|m{4cm}|m{5cm}}
\hline
    Method & Description & Typical papers & Advantages & Disadvantages    \\
\hline
Hashing & Hash the data record using locality sensitive hash function and add noise to each user's data point  &\tabincell{l}{\citep{nissim2017clustering},\\ \citep{stemmer2018differentially},\\\citep{stemmer2020locally}} & Only need a few round of iteration & Has a high noise \\
\hline
Bit vector & Map the data record into the Hamming space and perform the randomized response to each bit of the encoded vector & \citep{sun2019distributed} &Data dimension can be relatively reduced & Extra computation cost to make the distance consistent, and cannot achieve strict $\epsilon$-LDP definition\\
\hline
Transformation & Encode the data record as a binary string, and then perform the randomized response to the generated string and report the closest center differential privately &\citep{xia2020distributed} & The user's clustering information is protected & Not suitable for high dimensional data, and extra privacy consumption for cluster center selection\\
\hline
\end{tabular} \label{Usl}
\end{table*}

\subsection{Summary of statistical query with LDP}

We discussed three types of statistical queries, frequency estimation,  mean value estimation and range query. 
Frequency estimation is the most common query and aim at categorical data, we include the frequency estimation for the single attribute, multiple attributes and set-value data. 
Two main problems have to be considers  when estimate the item frequency, the effect of high dimensional data and statistical efficiency. 
Decreasing the data dimension and sampling are two important measures to deal with the high dimensional dataset. The principle of improving the efficiency is avoid going through all the items in the domain. 
Mean value estimation aims at the statistics for continuous numerical attribute. 
The research focus is on how to reduce the statistic variance as well. 
However, only query mean value is insufficient in many data analysis tasks. 
Researchers have to provide solutions answer more type of queries over numerical data, such as min and median. 
Range query is a generalized frequency estimation that queries the frequency of many points within the range.    
Hierarchy-based method is the core solution, whose essence is to reduce the data dimension to save privacy budget.

\section{Private learning with LDP} \label{mlldp}

Machine learning is an essential method of data analysis. 
Instead of answering simple statistical query, 
the aggregator has a clear data analysis target that can be achieved by training a corresponding machine learning model. 
According to the different algorithms, the user incorporates the LDP perturbation mechanism to the model training process, which is separated to supervised learning and unsupervised learning. 
Another group of research consider the learning process as an optimization problem, which is solved by defining a series of objective functions. We group these research as private learning in ERM.

\subsection{Supervised learning}

Supervised learning refers to the machine learning algorithms that learn a model mapping an input to an output based on a set of labelled training datasets. 
Naive Bayes classifier is one of the most simple and
effective supervised learning techniques in data mining. 
It is a probabilistic classifier that makes classifications using the maximum a posteriori decision rule in a Bayesian setting. 
Given a class label $c\in C$ and a conjunction of features $[x^1, \cdots x^d]$, 
the new instance is classified as the most probable label $c*$, where 
\begin{equation}
c*=arg\max \limits_{c\in C}\{P(c)\prod\limits_{i=1}^{d} P(x^i|c)\}
\end{equation}

$P(c)$ is distribution probability, $x^i$ is the value for $i$th attribute. 
Naive Bayes classifier is trained based on the statistics of conditional distribution of the feature and label. 
Naive Bayes classifier adopts attribute conditional independence assumption that the attributes are independent to each other. 

Two issues have to be considered to train a Naive Bayes classifier under local setting. 
The first issue is that the data dimension can be quite high. 
This issue is solved by partitioning the users into different groups, 
and users in each group report a pair of values (one attribute and the label). 
The second issue is that the correlation between the attributes and label needs to be preserved. 
Yilmaz et al. \citep{yilmaz2019locally} transform two attributes to one attribute, apply the frequency oracle to estimate the frequency and then calculate the conditional probability. 
Xue et al. \citep{xue2019joint} propose a similar idea as collecting key-value data that perturbs the attribute first and then perturbs the label value based on the attribute perturbation result. 
 
\textit{Discussion.} Simply partition the population to reduce the effect of high dimensional dataset is not optimal usually, as the population reporting the same attribute and label would be quite small with the increasing of the attribute dimension, which would reduce the statistical accuracy. 
Besides, Xue et al.'s method is essentially the same thing as Yilam et al.'s.  
They did not show the advantage of the method that perturb the label value based on the attribute perturbation result compared with the method apply frequency oracle directly.  
Assume both of them adopt GRR as the building block, they achieve the same statistical variance.
Table \ref{NBBCC} compares these two methods. 
There's still plenty of room to improve for Naive Bayes model training under local setting.

\subsection{Unsupervised learning}
Unsupervised learning refers to machine learning algorithms that get inferences from datasets with no pre-existing labels
As a typical unsupervised learning algorithm, 
clustering divides the unlabelled data record into a number of groups such that the data records are similar to each other in the same group. 
Specifically, given a set of data point $\{r_1, \cdots, r_n\}$, 
the output of clustering is a set of group center $\{c_1, \cdots, c_k\}$ and assigned points. 
One of the most fundamental clustering methods, k-means has been studied in the distributed setting recently. 

The basic k-means algorithm works as follows. It creates $k$ clusters by assigning each point to its closest center initially, and then re-calculates the center of each cluster iteratively until no changes happen. 
To protect the user's data locally, the user needs to report their perturbed data and closest center they belong to to the aggregator in each iteration. 
Two issues need to be considered. 
First, the high dimensional data causes too much noise added to each record. 
Second, the multi-round report weakens the privacy level. 
Currently, the research mainly focuses on reducing the data dimension while maintaining the distance property.  

Nissim and Stemmer\citep{nissim2017clustering} propose an algorithm for minimum enclosing ball, they make use of an LDP algorithm called GoodCenters, which works by hashing input points using a locality sensitive hash function to maximize the probability of a collision for close items, while minimizing the probability of collision for far items. Later Stemmer and Kaplan \citep{stemmer2018differentially} propose an improved algorithm for distributed K-means that significantly reduces the number of interaction rounds. Moreover, a new locally private k-means algorithm \citep{stemmer2020locally} is proposed recently, which further reduces the additive error in their work. 
Sun et al. \citep{sun2019distributed} use Bit Vector mechanism to encode the user data into Hamming space, which eliminates the semantic information while preserving the initial distance between records. 
Then the traditional randomized response is performed on each bit of the data to ensure the indistinguishability. 
Xia et al. \citep{xia2020distributed} propose a feature transformation method that encodes the user's data as a production of a binary string with a coefficient, which provides a trade-off between accuracy, communication costs, and privacy. To enhance the privacy and hide the cluster each user belongs to, the user's closest center is perturbed by the LDP protocol as well. 
Table \ref{Usl} compares these methods in regard to different encoding mechanisms. 

\textit{Discussion.} The mean value of data in each dimension is usually calculated as the new center in each cluster. 
However, the population can affect the accuracy of the calculation significantly. 
The statistics for clusters with a small population would have a big variance. 
The strategy to solve this problem needs to be explored in the future. 
Besides, though some researchers \citep{stemmer2018differentially} try to reduce the iteration round, 
reducing the privacy consumption in distributed clustering remains a challenge.

\begin{table*}[htpb]  \centering
\begin{threeparttable}[b]
  \caption{Private learning risk bound}\label{riskbound}
  \begin{tabular}{m{1cm}|m{1.5cm}|m{1.8cm}|m{2cm}|m{2cm}|m{5cm}|m{1cm}}
\hline
    Paper &  Model & Learning algorithm & Assumption for loss function  & Perturbation method  & Risk bound/Sample complexity  & Privacy level \\
\hline
\citep{smith2017interaction} & \tabincell{l}{Non-\\interactive} &--\tnote{1}  & Convex, 1-Lipschitz &  Gradient 
& Exponentially depend on $d$ & $\epsilon$ \\
\hline
\multirow{3}*{\citep{zheng2017collect}} & \multirow{3}*{\tabincell{l}{Non-\\interactive}}  & Sparse linear regression & Convex  & Input  &Logarithmically depend on $d$ & $(\epsilon, \delta)$ \\
\cline{3-7} & & Kernel ridge regression & Convex, Lipschitz & Input &Polynomial dependences on $d$& $(\epsilon, \delta)$ \\
\cline{3-7} & &-- & Convex, smooth generalized linear function & Input   &-/Quasi-polynomial with respect to $\frac{1}{\alpha}$ & $(\epsilon, \delta)$ \\
\hline
\multirow{2}*{\citep{wang2018empirical}} & \multirow{2}*{\tabincell{l}{Non-\\interactive}} & --& $(\infty, T)$-smooth & Objective & 
-/Polynomially dependent on $\frac{1}{\alpha}$ for constant or low dimensional case & $\epsilon$\\
\cline{3-7}& & --&Convex, generalized linear function, 1-Lipschitz & Objective  & depend on $n$ and Gaussian width of the constrained set & $\epsilon$\\
\hline
\multirow{3}*{\citep{wang2018high}, \citep{wang2019sparse}} & \tabincell{l}{Non-\\interactive } & Sparse linear regression & -- & Input  &  Depends on $d\log d$& $\epsilon$ \\
\cline{2-7} & Interactive & Sparse linear regression &-- & Gradient & Depends on $\log d$ for low dimensional case & $\epsilon$ \\
\cline{2-7} & Interactive & Sparse linear regression & --& Input & Depends on $\log d$ if only the lables are required to be private & $(\epsilon, \delta)$ \\
\hline

\end{tabular} \label{NBCC}
\begin{tablenotes}
\item[1] --: Not specified
  
\end{tablenotes}
\end{threeparttable}
\end{table*}

\subsection{Private learning in ERM}

The purpose of private learning framework is to design a private learner that outputs an approximately accurate model and preserve the differential privacy of the training samples. 
Empirical risk minimization (ERM) is a typical technique used to select the optimal model from a set of hypotheses. 

Given a dataset $D=\{r_1, \cdots, r_n\}$, hypothesis $h\in \mathbf{H}$ and loss function $\ell(h(\mathbf{w},r_i),y_i)$, 
the goal of empirical risk minimization is to identify the $\mathbf{w}$ that minimizes the empirical risk $R_n(\mathbf{w})$ on dataset $D$ shown in Eq. \ref{erm}. 
\begin{equation}\label{erm}
R_n(\mathbf{w})=arg \min \limits_{h\in \mathbf{H}} \frac{1}{n} \sum \limits_{i=1}^n \ell(h(\mathbf{w},r_i), y_i)
\end{equation}
 
The utility of private ERM is measured by the difference between the real risk $R_n(\mathbf{w})$ and private risk $\hat{R_n}(\mathbf{w})$, defined as a risk bound, or the sample complexity that the number of samples are needed to achieve a bounded accuracy. Both of which can be considered as implementation of Eq. \ref{error}. 
\begin{equation} \label{error}
Pr[|\mathcal{M}(D)-\hat{\mathcal{M}}(D)|\leq \alpha]>1-\beta
\end{equation}

ERM can be implemented for certain learning tasks by choosing different loss functions. 
For example, linear regression defined the loss function as maximum likelihood estimation, while logistic regression defined the loss function as the logistic loss. 
If the task is simple statistical analysis, such as mean or median statistic, the loss function is simplified to $\ell (\theta, r_i)$, where $\theta$ is the estimated value. ERM is used to minimize the empirical risk $\hat{L}=\frac{1}{n} \sum \limits_{i=1}^n \ell (\theta, r_i)$. 
To make the ERM solution tractable, the loss function is usually assumed to be convex, therefore, minimize the empirical risk can be considered as a convex optimization problem. 

ERM under local differential privacy includes two kinds of protocols, interactive and non-interactive.  
The interactive model allows the aggregator to collect the data sequentially, the reported value of $u_i$ is based on user $u_i$'s true value and user $u_{i-1}$ perturbed value. The non-interactive model requires the data to be collected at once. Currently, the majority of the work of statistical analysis uses the non-interactive model. Three methods are considered to incorporate differential privacy into the learning process: input perturbation, gradient perturbation, and objective perturbation. The input perturbation inserts the noise into the data record directly. The gradient perturbation adds the noise to the gradient generated by the gradient descent optimization algorithm, while objective perturbation adds noise to the objective function prior to learning.  

Kasiviswanathan et al. \citep{kasiviswanathan2011can} initiate the study of private local learning under the interactive model. 
Later, Feldman et al. \citep{feldman2017statistical} propose a wide range of convex ERM problems for statistical query and  Duchi et al. \citep{duchi2013local, duchi2018minimax} give optimal upper and lower bounds.   
Paper \citep{zheng2017collect, smith2017interaction, wang2018empirical} study the convex optimization problem under the non-interactive model.  
Smith et al. \citep{smith2017interaction} focus on convex Lipschitz functions and presented a result that the sample size is exponentially dependent on the dimensionality $d$. 
Wang et al. \citep{wang2018empirical} show that in the case of constant or low dimensions, if the loss function is $(\infty, T)$-smooth, the exponential dependency can be avoided. In the high dimensional case, if the loss function is a convex generalized linear function, the error can be bounded by the Gaussian Width of $\mathcal{C}$ and $n$ instead of $d$. 
Zhang et al. \citep{zheng2017collect} study some specific loss functions under high-dimensional space, such as spare linear regression and kernel ridge regression, and prove the polynomial dependence of excess risk or square error over $\log d$ and $\frac{1}{n}$. 
Wang et al. \citep{wang2018high, wang2019sparse} study Empirical Risk Minimization problem with sparsity constraints, and achieved an upper bound that
depends only logarithmically on $d$. 
Van et al. \citep{van2019user} extend the local differential privacy framework in unconstrained online convex optimization by allowing the provider of the data to choose their privacy guarantees.

\textit{Discussion.} The risk bounds of ERM are highly associated with the dimension of the data and the size of the training samples. Though researchers try to relax the dependency by both adding extra assumptions and relaxation of the privacy level, the polynomial dependency seems still cannot be avoided. 
Besides, the constraints to the loss function might hinder the practical development of the private learning framework. We summarise the risk bounds of private learning algorithms in Table \ref{riskbound}.

\begin{table*}[htpb]  \centering
  \caption{Practical Deployment}\label{tab-parameter}
  \begin{threeparttable}
  \begin{tabular}{m{1.2cm}|m{1.6cm}|m{3cm}|m{2cm}|m{1.5cm}|m{2cm}|m{3cm}|m{0.6cm}}
\hline
    Company & Deployment & Purpose/Functionality & Techniques & Population & Parameters & Limitations & Open source  \\
\hline
Google & Chrome \tabincell{l}{Browser\\ (2014)}& Collect up-to-date statistics about the activity of their users and their client-side software & \tabincell{l}{2-level RR\\ memoization \\Bloom filter} & 14 million & \tabincell{l}{ $\epsilon=0.5343$\\ $h\tnote{1}=2$ \\$k\tnote{2}=128$} &
Not suitable for data with frequent changes & Yes \\
\hline
Apple & \tabincell{l}{macO\\iOS10 (2016)}& Estimate the frequencies of elements& \tabincell{l}{RR\\ CMS \\HT\tnote{3}}&  Hundreds of millions &  \tabincell{l}{$\epsilon=2\sim 8$\\$m\tnote{4}=256\sim 32768$\\$h=1024\sim 65536$} & The overall privacy cost for each device is unbounded & No\\
\hline
Microsoft & \tabincell{l}{Windows 10\\ (2017)} & Repeated collection of counter data \tabincell{l}{mean estimation \\ histogram estimation}&\tabincell{l}{1BitMean\\ dBitFlip\\$\alpha$-point rounding\\memoization}  & millions & $\epsilon=1$ & Not suitable for data with significant changes & No\\
\hline
SAP & \tabincell{l}{HANA 2.0 \\SPS03 (2018)} & \tabincell{l}{ Count\\Sum\\Average} & LM\tnote{5} & -- & Leave it up to the data consumer & Only support numerical value The added noise is unbounded & No \\

\hline
\end{tabular}
\begin{tablenotes}
  \item[1] $h$ number of hash functions 
  \item[2] $k$ Bloom filter size
  \item[4] $m$ CMS size
  \item[3]  $HT$ Hadamard transform
  \item[5]  $LM$Laplace mechanism
\end{tablenotes}
\end{threeparttable}
\end{table*}

\subsection{Summary of private learning with LDP}

Machine learning is an essential method of data analysis used to design a model to learn the trend and dependence between attributes and target variable. 
Due to the high dimensional user data and many rounds of iteration, 
it's difficult to train an accurate model with a high level of privacy guarantee. 
Distributed environment makes it even more challenging. 

Naive Bayes classification and decision tree are two popular and simple supervised learning algorithms. 
Besides reducing the added noise, the correlation between the attributes and label needs to be preserved. 
Currently, partition users or attributes to save privacy budget is the only way in the literature to enhance the model accuracy for Naive Bayes classification. 
Decision tree learning has not been studied in the local model yet. 

Clustering is a typical unsupervised learning algorithm. 
Users need to cooperate with the aggregator to update the centers iteratively. 
Though several works try to reduce the iterative round to save privacy budget, it is challenging to reduce it to one round. 
Another effective way to enhance the accuracy is to reduce the data dimension. 
The essential challenge for this method is how to maintain the distance between records while doing the dimension reduction. 
Several methods have been proposed but there is still room for further improvement, 
especially for the process of the protection for the data center that the data record belongs to. 

Private learning combines differential privacy and various learning algorithms to train a model while preserving the training samples. 
ERM helps to select the optimal model by transferring the learning problem to convex optimization problem. 
PAC theory builds the relationship between the sample complexity and model accuracy, 
which is highly related to the data dimension. 
Researchers are trying to relax the dependency on data dimension and narrow the sample complexity gap between the interactive and non-interactive model by adding the extra constraint to the object function, which makes it impractical for real applications.


\section{Applications of local differential privacy} \label{aldp}

\subsection{Applications in practice} 
Google proposed RAPPOR in 2014, and deployed it to the Chrome Web browser to collect the clients' preferences (e.g. default home page and search engine), which is the first internet-scale deployment of LDP. As aforementioned, RAPPOR hashes the item into a short bit vector and performs the randomized response to each bit of the vector. To relieve the collision problem, they utilize Bloom filter and cohort methods to introduce the redundancy to improve the false positive rate. The decoding process of RAPPOR is quite complicated, which requires sophisticated statistical techniques and inefficient. 
RAPPOR supports multiple collections of the data from the same respondent over time. 
The idea is memoization that instead of directly reporting the perturbed value $B'$ on every request, the client reports a randomized version of $B'$. 
The advantage is that it provides longitudinal privacy, but it is only in cases when the user's value does not change or changes in an uncorrelated fashion.    

Apple launched LDP for the first time in macOS Sierra and iOS 10 to protect the user's activity while providing the QuickType and emoji suggestions.  
The general idea is that the users' data with the varying size is encoded as matrix of fixed size using Count Mean Sketch (CMS) technique. Then the Hadamard basis transformation is applied to the hashed encoding. 
Instead of sending the entire row to the server, only a single bit is sampled at random to send to the server. This reduces the communication cost at the expense of some accuracy. 
The deployed LDP algorithm does not consider the privacy cost for repeat collection. 
Instead, they set a limit on the number of records that can be collected for each use case. 
To trade-off utility with privacy budget, device bandwidth, and server computation cost, the parameters are selected based on the effect of them to the variance of the estimated counts.   

Microsoft deployed LDP to Windows 10 from 2017 to collect the number of seconds that a user has spent using a particular app. 
The proposed privacy framework is similar to RAPPOR. 
Memoization is also used for continual collection. 
The difference is that $\alpha$-point rounding mechanism is proposed, which enable slight changes in user's data without violation of predefined privacy. 
The key idea is to discretize the domain, then the values stay within a single segment trigger the same memorized response. 
But it is not working for the situation that the value changes significantly.  

SAP integrates privacy-enhancing techniques into database management system SAP HANA since 2018. 
Different from other privacy-enhanced database management platform that adopts centralized differential privacy, such as PINQ \citep{mcsherry2009privacy}, SAP adopts local differential privacy to avoid the hassle and overhead of maintaining a privacy budget. 
The general idea is that transform the dataset by adding Laplace noise to each record of the specified column of the dataset before performing any query. 
The system utilizes seed to make the noisy result reproducible if no changes have occurred to the database. 
The content of privacy view in the system is always recalculated to check whether or not any changes result in violations of the current privacy definition. 
The privacy enhanced system let the data consumer set the privacy parameter. 
SAP HANA is claimed to be the first data management production that addresses data privacy through an integrated and domain independent approach. 
It provides an easy-to-use solution, but we see the potential for further improvements. 
For example, the noise added to the database is unbounded, the solution is only suitable for numerical attributes, and the supported query types are limited.  
   
\textit{Discussion.} 
The practical deployment of LDP currently is limited to the count and average functions. 
Deploying LDP in practice is quite challenging. 
First, 
there is limited experience on how to determine the privacy level. 
Since the utility requirements predominately depend on the application and the selection of the privacy level is based on the privacy characteristics of the underlying dataset for each use case. 
How can we well balance the privacy and utility? Current practice is to choose the parameter according to the limited experience or run a number of simulations over a small set of datasets to find how the parameters affect the performance to make the selection. The general range of $\epsilon$ is $[0.1\sim 10]$. 
Second, the privacy level is unbounded due to the continuous collection. 
Because of the sequential combination property of differential privacy, 
the privacy level becomes weaker and weaker as the increasing of the number of query times. 
Memoization is the only way to prevent such averaging attack. 
However, the drawback is that the user's data cannot be changed or change significantly, which makes continuous statistic meaningless.

\subsection{Applications in various domain}

\subsubsection{Federated learning} 

Federated learning, which proposed by Google in 2017 \citep{mcmahan2017federated}, is a recent advance in privacy-preserving machine learning, where the model enables the mobile phone to learn from its local data and only the parameters are sent to the server for global model training. 
Distributed stochastic gradient descent (SGD) is commonly adopted in federated learning for training machine learning models. 

Though the users' data do not leave their own devices, the adversaries can use differential attacks to determine which mobile users have been included during the learning process \citep{smith2017federated,lyu2020threats} and the novel discrimination on client identity enables the generator to recover user specified private data \citep{wang2019beyond}. For example, Nesterov \citep{nesterov2013introductory} demonstrate that the model-inversion attack can recover the images in the facial recognition system and Shokri et al. \citep{shokri2017membership} show the vulnerability of deep learning models trained on sensitive data for membership inference even when the sensitive data are released as black box models.

To prevent the attack from the adversaries, algorithms using differential privacy for distributed learning have been proposed \citep{mcmahan2017learning,geyer2017differentially, triastcyn2019federated, wei2019federated,choudhury2019differential}. The main idea is to add the calibrated Gaussian noise to the parameter reported to the server at the user side to balance privacy and predictive accuracy. 
The majority of the work only protect the record-level privacy. 
However, McMahan et al. \citep{mcmahan2017learning} claim that for problems like language modelling, protecting the single record is insufficient. 
Because the sensitive word maybe typed several time, all of which should be protected. 
Therefore, they proposed to add user-level privacy (LDP) protection to the federated averaging algorithms. 
Besides adding Laplace or Gaussian noise, randomized response is used in private federated learning as well. 
For example, Wang et al. \citep{wang2019collecting, zhao2020local} utilize the proposed mean value perturbation mechanism to perturb the gradient to ensure the unbiased mean value estimation under federated learning framework.  
Chamikara et al. \citep{arachchige2019local} apply randomized response to local deep learning training. Specifically, they divide the structure of a convolutional neural network into two main modules, convolutional module and ANN module, and introduce an intermediate randomization layer which is named as LATENT in between these two modules. LATENT converts the input values to binary values and then performs the randomized response. 

\textit{Discussion.} LDP provides much stronger privacy protection for users during training process. 
While McMahan et al. show that the model trained with strong privacy guarantees (user-level privacy), showing no significant decrease in model accuracy given a large enough dataset. 
The problem is that does LDP provides too strong privacy in practice at the cost of other resource consumption. 
Besides, some unique issues of federate learning model need to be considered when combining with privacy-preserving technology, such as the communication efficiency and power limitation of mobile devices. 
Because more round of iterations may need to reach convergence due to the involvement of differential privacy.

\subsubsection{Reinforcement learning}

Reinforcement Learning is a type of machine learning technique that enables an agent to learn in an interactive environment. 
It has been well adopted in artificial intelligence (AI) as a way of directing unsupervised machine learning through rewards and penalties in a given environment. 
While the environment may be related to some private information, such as the private indoor layout. 
Pan et al. \citep{pan2019you} show the vulnerability of reinforcement learning to potential privacy-leaking attacks, and they recover the map structure successfully. 
  
Ono et al. \citep{ono2020locally} propose a local differential privacy algorithm based on asynchronous advantage actor-critic (A3C) for distributed reinforcement learning to obtain a robust policy. 
They propose a Laplace method and a random projection method to introduce the randomness to the distributed gradient that satisfies LDP to prevent information disclosure. 
Gajane et al. \citep{gajane2018corrupt} initiate the study of the LDP multi-armed bandit problems and proposed an LDP bandit algorithm to hide the reward, which is considered refer to the users' activates that involve private information. 
Basu et al. \citep{basu2019differential} provide a unified framework to prove minimax lower bounds on the regret of both differentially private multi-armed bandits, and they show that when differential privacy is achieved using a local mechanism, the regret scales as a multiplicative factor of $\epsilon$. 
Latter, Ren et al. \citep{ren2020multi} prove a much tighter regret low bound and develop corresponding LDP upper confidence bound algorithm by adding Laplace noise to the reward or converting the rewards to Bernoulli responses. 

\textit{Discussion.} Artificial intelligence has attracted a large amount of attention in recent years. 
Zhu and Yu \citep{zhu2019applying} explored the possibility of applying differential privacy mechanism in AI. 
And we also see the potential of LDP, which comply with the setting of multi-agent system and distributed learning quite well. 
The application of LDP in this area is at a very early stage. 
More research is needed to help us to understand the functionality and effect of LDP in reinforcement learning.   

\subsubsection{Social network}
Online social networks provide an unprecedented opportunity for researchers to analysis various social phenomena. These network data is normally represented as graphs, which contain many sensitive individual information \citep{zhu2018iteration}. To protect the privacy, a synthetic social graph is released instead of the original one.   
The synthetic graph generation algorithm includes Erdos-Renyi \citep{erdos1959random}, Chung-Lu \citep{aiello2000random}, Kronecker \citep{leskovec2010kronecker}, BTER \citep{seshadhri2012community} and so on, which takes some graph statistical information as input, such as node degrees,
and some high-level structural properties of the input graph. 
Besides the high dimensional problem, two additional challenges have to be considered when developing the solutions to protect the user's privacy under local differential privacy protection. 
\begin{itemize}
\item First, the current study of local differential privacy mainly focus on some simple statistic, such as count and mean. While more detailed information, such as edge-level information, are needed to fit the graph generation models.
\item Second, each user only has a local view of the group, it is difficult to obtain the global information of the entire graph, such as the submatrix of the adjacency matrix in Kronecker Graph model. 
\end{itemize}
 
Therefore, how to reduce the impact of high dimensional data and how to utilize simple statistics to obtain more graph information are two major research questions under local differential privacy protection. 
Qin et al. \citep{qin2017generating} partition the users into disjoint groups, where similar users are grouped together. 
They obtain more structural information from each user by building node-to-group connectivity. Instead of collecting the statistic information (the number of connections) for each user. 
Zhang et al. \citep{zhang2018two} let the user report a random neighbour list. To defend the drawback of applying randomized response directly, they partition the users into disjoint groups according to their similarity as well. 
Only the nodes in the same group are included in the neighbour list, which significantly narrows down the range of the randomized response. 
Besides, Sun et al. \citep{sun2019analyzing} argue that the traditional local differential privacy definition can only protect the user, but not the user's neighbours. To solve this problem, they proposed a new concept, Decentralized Differential Privacy (DPP), to protect all graph participants. 

\textit{Discussion.} The essential challenge for graph publication under local model is to get as much structural information as possible with the limited view. Currently, only edge local differential privacy is considered under local model. 
Node privacy is quite challenging in centralized model, the distributed environment makes it more difficult. 

\subsubsection{Location privacy}
GPS-enabled devices allow the location information to be easily collected and provide opportunities for the development of location-based services, such as tracking system, social network services, and location-based advertising. 
Not alone the location information, the user's home address, his religious practice, behaviour, and habits can be disclosed.  
LDP has been used to collect the user's location data and statistic the population distribution. 
The location data is treated as a categorical attribute instead of numerical value. 
Similar to the general high-dimensional categorical attribute, location data has a large domain size. 
Another character of location data is that the location is easy to be generalized to a larger region. 
The statistics of the population in a specific region is a typical range query. Therefore, the hierarchy tree is used as a common tool to generalize the location and reduce the statistical variance. 

Let the root node be the highest level of the tree, which includes all the locations of the universe. 
Zhao et al. \citep{zhao2019ldpart} iteratively split the node into disjoint sub-partitions with more detailed representations in a top-down manner based on the noisy count of the node. And make use of the noisy count of each leaf node to construct the synthetic dataset. 
Chen et al. \citep{chen2016private} allow the user to choose a safe region (the internal node in the tree) to report, which has a much lower statistical variance compared with reporting the location in the lowest level. 
Kim et al. \citep{kim2019workload} consider the workload-aware indoor positioning data collection. 
They let the user report a binary vector, which includes part of the path sequence, to statistic the noisy count of all the nodes in the tree. 

\textit{Discussion.} Even though previous works provide various solutions to location privacy, they mainly focus on the population statistics, location privacy issues in other applications have not been considered under local differential privacy protection. For example, providing recommendations about nearby points of interest without disclosing the user's location and allowing effective task assignments in the spatial crowdsourcing system. 

\subsubsection{Recommendation system}
Recommendation system is one of the most popular applications in e-commerce and online social networks \citep{zhu2017differentially}. It predicts the ratings or preferences a user would give to an item utilizing the user's past behaviour as well as similar decisions made by other users. However, continual observation of the recommendations with some background information enables an adversary to infer the individual's rating or purchase history \citep{calandrino2011you}. 

Matrix factorization is a class of collaborative filtering algorithms, which is one of the most widely used techniques in recommendation system. 
Each user $i$ is characterized by a profile vector $u_i\in \mathcal{R}^d, 1\leq i \leq n$ and each item $j$ is characterized by a profile vector $v_j\in \mathcal{R}^d, 1\leq j\leq m$. 
User $i$'s rating of item $j$, which is denoted by $r_{ij}$, 
is then approximated by the inner product of $u_i$ and $v_j$. 
That is $r_{ij}\doteq u_i^{\mathrm{T}}v_j$. 
Matrix factorization computes the user profiles $u_i$ and the item profiles $v_j$ by minimizing the regularized mean squared error on the set of known ratings. 
While gradient descent is one of the popular approaches to solve the nonconvex optimization problem. 
It iteratively learns the $u_i$ and $v_i$ based on some updating rules. 
As profile vector $u_i$ only relies on user $i$'s own ratings, the user profiles $u_i$ can be computed in user's devices without reporting any data to the recommender when the $v_i$ are shared by the recommender. 
While the $v_i$ can be updated by the aggregator by collecting the gradients from users. 
Therefore, the user can protect their data by perturbing the gradient before sending it to the server to satisfy local differential privacy. 

Shin et al. \citep{shin2018privacy} protect the items and ratings by using randomized response and Harmony respectively. Jiang et al. \citep{jiang2019towards} apply two times randomized response as RAPPOR to prevent the rated items from being disclosed and adding Gaussian noise to hide the rating. 
As the number of items is very big, and the calculation takes multiple iterations, applying the general randomized response to each value results in unacceptable estimation error. 
Shin et al. \citep{shin2018privacy} utilize dimension reduction and sampling methods to improve the recommendation accuracy. 
Besides, many other works \citep{guo2019locally, xue2019distributed, DBLP:conf/ijcai/HuaXZ15, shen2016epicrec, shen2014privacy} try to prevent the user's data from being disclosed to the untrusted server as well. They protect the user's ratings \citep{DBLP:conf/ijcai/HuaXZ15} or items \citep{shen2016epicrec, shen2014privacy} by adding Laplace noise, which provides privacy for a single rating or item instead of the whole data record. 

\textit{Discussion.} Rating dataset is high sparse. In practice, the recommendation based on the original dataset is not very accurate due to the missing values. 
However, some randomized algorithms can improve the accuracy of the recommendation. 
For example, Yang et al. \citep{yang2017privacy} utilize Johnson-lindenstrauss transform to get a much accurate recommendation even compared with non-privacy method in the centralized model. 
Therefore, such randomized methods might be used to reduce the impact of the large noised added under the local model. Not only because filled missing value, but also the reduced dimension.

\subsection{Summary}

Local differential privacy has not been studied extensively compared with centralized differential privacy. 
Part of the reason for this is that there are intrinsic limitations in what one can do in the local model. 
Currently, local differential privacy mainly focuses on simple statistics, such as frequency estimation, mean value computation, and applications based on these simple statistics. 
Though it is rarely mentioned, the simple private statistic can be potentially applied to various application areas, such as smart meters \citep{ou2020singular,lyu2017privacy,lyu2018ppfa}, smart intelligent transportation system, crowdsourcing and medical data analysis \citep{kim2018privacy}. 
 
In addition, other applications are also proposed in the literature. 
For example, 
Yang et al. \citep{yang2019collecting} study how to collect the users' preference ranking under local DP. 
Sun et al. \citep{sun2018truth} study the truth inference problem on sparse crowdsourcing data.  
Private Principal Component Analysis (PCA) problem has also been studied under both interactive \citep{balcan2016improved, ge2018minimax} and non-interactive \citep{wang2020principal} local model. 
Recently, Choi et al. \citep{choi2018guaranteeing} introduce the application of local differential privacy on Ultra-low power system that supports low resolution and fixed point hardware. Lyu et al. \citep{lyu2019fog,lyu2020foreseen,lyu2020towards} introduce different LDP protocols to preserve the extracted hidden representation in deep inference and NLP domain. 
Paper \citep{ding2018comparing, gaboardi2017local, gaboardi2018locally, sheffet2018locally} study hypothesis testing under the local setting. 

Local differential privacy still has much unknown potential, all these mentioned research is a good starting point for extending the application of local differential privacy in the future.

\section{Research gaps and research direction} \label{rgrd}

Local differential privacy is a relatively new research filed. 
Due to its unique properties, LDP has many challenging problems need to be solved. 
We identify the research gaps for LDP and propose a few research directions in this section. 

\subsection{Research gaps}
There are many problems that have not been well studied under local differential privacy. 

First, as aforementioned, LDP perturbs every data record, 
which reduces the statistical accuracy significantly compared with centralized model. 
The higher dimensional dataset makes it even worse. 
There are quite a lot of research focus on high dimensional data statistical, 
the solution includes partition, hash, and matrix transformation. 
However, the majority of exiting solutions are focus on the single attribute. 
For the multiple attributes statistics, the general way is to sample one or few attributes to report, which inevitably reduces the utility. 
Therefore, there is still a long way to go for high dimensional data statistical over multiple attributes. 

Second, the principle of LDP getting accurate statistics is that the added positive and negative noises can be cancelled out and the perturbation mechanism is usually designed based on the predefined query. 
The bias of the perturbed data is usually very big. 
The aggregator is hardly to perform other types of queries on the collected data. 
That is, the data perturbation method is bounded with the predefined query, the aggregator can only get the accurate estimation for the specified query. 
However, supporting multiple types of query is the fundamental requirement of data analysis, how to make it achievable and ensure accurate estimation is quite challenging under local environment. 

Third, users' data change over time and the statistics change accordingly, which needs to be re-computed periodically in practical. 
Continuous data statistics have been studied well under centralized differential privacy model, but has not been well studied with local differential privacy protection. 
Currently, memoization has been proposed in the literature.  
However, it is not working for data with frequent changes \citep{erlingsson2014rappor} or data change significantly \citep{ding2017collecting}. 
Besides, only Joseph et al. \citep{joseph2018local} consider the continuous statistic. 
But, the proposed voting method is only applicable for tracking statistical problems for evolving data that the stable number of users receive new information in each round. 
The problem of statistic stream data needs to be further explored 
and more practical scenario that both the number of participated users and user's value are changed need to be considered when developing the solutions.

\subsection{Research directions}
Besides the research to solve the aforementioned gaps, we identify a few new research directions shown as follows. 

\subsubsection{Relaxation of LDP}
Local differential privacy provides a very strong privacy guarantee for any pair of data record. 
However, in many applications, some items are not very sensitive for the user in the data domain.   
For example, in the geolocation application, the user may do not care to disclose the city he lived in, 
but hope to hide the precise location where he located at. 
Therefore, the locations outside the city he lived in are not sensitive to him. Similarly, when collecting website visits, the website type is less sensitive than the particular website. All these examples motivate the relaxations that incorporate the application's context into the privacy definition.  

Acharya et al. \citep{acharya2019context} propose block structured LDP (BSLDP), which preserves the privacy of items that are close to each other. Specifically, the model divides the data domain into various partitions, the element pairs in the same partition and those in different partition have different levels of indistinguishability. 
Gursoy et al. \citep{gursoy2019secure} and Alvim et al.\citep{alvim2018local} propose metric-based LDP. 
Metric LDP is a direct extension of $d_{\chi}$-privacy \citep{chatzikokolakis2013broadening} in centralized DP, which controls the indistinguishability by items' distance $d(·, ·)$ in addition to the privacy budget. 
Later, Xiang et al. \citep{xiang2019linear} apply the Metric DP to more analytical tasks, such as linear counting query and mult-dimensional range query, which achieve significant gains in utility.  

Currently, the relaxation of LDP is mainly aimed at location dataset. 
More works need to be done for various data types and application scenarios. 
First, how should we define the privacy that develops the new definition, which includes identifying the different privacy concern and how to quantify privacy level. 
Second, how to design the randomized algorithm to achieve the new definition and analyse the sample complexity are also challenging.

\subsubsection{Privacy amplification}
Privacy amplification refers to the strategy that enhances the privacy level without or with a very little effect on the data utility. 
It enables the algorithm to achieve same privacy level with much smaller samples or achieve higher accuracy with same privacy level. 
Due to a large amount of noise and poor utility of LDP (especially compared to centralized DP), 
the privacy amplification techniques provide possible solutions for the practical application of LDP. 

Shuffling is a typical method, which has been investigated in the literature \citep{cheu2019distributed,balle2019privacy}. 
It achieves better utility by relaxing the LDP trust model. 
The general idea is to insert a shuffler between users and the server to break the linkage between the report and the user identification. 
Erlingsson et al. \citep{erlingsson2019amplification} prove that if the users send the data with $\epsilon_b$-LDP protection, the shuffling step amplifies the privacy guarantee to be $(\epsilon_a, \delta)$-DP, where $\epsilon_a=O(\epsilon_b/\sqrt{n})$. The shortcoming of shuffling model is that if the auxiliary server colludes with the server, 
the privacy falls back to the original LDP model, the privacy guarantee is completely broken. 
Distributed Differential Privacy (DDP) combined with cryptosystem can be utilized to solve this problem, as evidenced by~\citep{dwork2006our,lyu2017privacy,lyu2019distributed}. The notion of DDP reflects the fact that the noise in the target statistic is sourced from multiple parties. Approaches to DDP that implement an overall additive noise mechanism by summing the same mechanism run at each party (typically with less noise) necessitates mechanisms with stable distributions (like Gaussian distribution, Binomial distribution)---to guarantee proper calibration of known end-to-end response distribution---and cryptography for hiding all but the final result from participants~\citep{lyu2020lightweight}.
A more recent work by Wang et al. \citep{wang2019practical} introduces multiple auxiliary servers to make this threat more difficult. As long as the server cannot collude with all the auxiliary servers, there is still some privacy amplification effect, but this introduces more communication costs. 

This line of research proves that privacy level of LDP can be affected by extra data processing. 
Though at early stage, quantified privacy level for more privacy amplification techniques need to be explored for LDP, such as sub-sampling \citep{balle2020privacy}, iteration \citep{feldman2018privacy, asoodeh2020privacy} and diffusion mechanism \citep{balle2019privacy}. 

\subsubsection{Solutions for small population}
Due to large amount of noise added to the whole dataset, LDP needs a large number of participants to cancel out the noise to ensure the statistical accuracy. 
For example, both Google and Apple's deployment collects one dozen million samples. 
However, small-scale enterprises may not able to collect such a big volume of samples, 
and the populations can be small for some specific applications, such as collect the people's distribution at a certain place. 
Moreover, even the sample is large, the effective sample size can be reduced in the high privacy regime. 
For example, Duchi et al. \citep{duchi2013local} state that when $\epsilon \in [0, \frac{22}{35}]$, $\epsilon$-local differential privacy reduces the effective sample size from $n$ to $4\epsilon^2n$. 

Some works \citep{murakami2018toward, gursoy2019secure} try to reduce the sample size by improve the statistical estimation accuracy. 
However, the effect is limited. 
A possible solution is that combine with other privacy techniques to protect the user's data, which enable less noise added to each data record without disclosing the privacy. 
While other problems may introduced for the hybrid method, such as computational and communication cost might increase. 
Besides, some hybrid strategies \citep{avent2017blender} that enable some users report their true values to reduce amount of add noise can be explored. The intuition is that people have different privacy preferences.  
We can save a lot of privacy budget from users who are not care their information or trust the curator. 
It's requisite and challenging to develop specific solutions for applications with a limited number of participants and fill the gaps in the theoretical analysis.

\section{Conclusion} \label{con}

Big Data provides a tremendous amount of detailed data for organizations to improve decision making. 
However, these data are highly sensitive, and the leakage of these data has received great focus.  
Localized differential privacy, as a new privacy protection model after centralized differential privacy, breaks the assumption that the curator has to be trusted by perturbing the data before leaving users' devices. 

This paper presents a comprehensive survey of work on local differential privacy including an overview of the huge amount of literature in two major local differential privacy research streams: the statistic query and private learning. 
We identify three general statistic queries, which includes frequency, mean, and range. 
We discussed the existing methods answering these queries, 
and further analyzed and compared the typical methods and techniques, 
which provide a comparative review for further research.  
We classify the research about private learning into supervised learning, unsupervised learning and private learning in ERM. 
We discuss and analyze the research in each category. 
In addition, we explored the application and practical deployment of local differential privacy and proposed
future research directions based on current research status. 

The local differential privacy technology is an emerging research field,
the research of it is incomplete, and it still has much unknown potential. 
The survey of literature in this paper provides an overview of existing works and is intended as a starting point for exploring new challenges in the future.

 \section*{Acknowledgment}
  This research is supported by the National Research Foundation, Prime Minister's Office, Singapore under its Strategic Capability Research Centres Funding Initiative

%
%
%
%




%

\appendices




\ifCLASSOPTIONcaptionsoff
  \newpage
\fi



%

\bibliographystyle{IEEEtran}
\bibliography{IEEEabrv,references}

\end{document}